\documentclass[aps,twocolumn,superscriptaddress,showpacs,showkeys]{revtex4-1}
\usepackage[]{graphicx,color}

\usepackage[utf8]{inputenc}
\usepackage{CJKutf8}
\usepackage{xspace}

\definecolor{darkgreen}{rgb}{0,0.5,0}

\begin{document}
\title{First-principles investigation of Nd(Fe,M)$_{12}$ (M = K--Br) and 
Nd(Fe,Cr,Co,Ni,Ge,As)$_{12}$: 
Possible enhancers of Curie temperature for NdFe$_{12}$ magnetic compounds}

\author{Taro \surname{Fukazawa}}
\email[E-mail: ]{taro.fukazawa@aist.go.jp}
\affiliation{CD-FMat, National Institute of Advanced Industrial Science
and Technology, Tsukuba, Ibaraki 305-8568, Japan}

\author{Hisazumi \surname{Akai}}
\affiliation{The Institute for Solid State Physics, The University of Tokyo,
Kashiwa, Chiba 277-8581, Japan}
\affiliation{ESICMM, National Institute for Materials Science,
Tsukuba, Ibaraki 305-0047, Japan}

\author{Yosuke \surname{Harashima}}
\affiliation{CD-FMat, National Institute of Advanced Industrial Science
and Technology, Tsukuba, Ibaraki 305-8568, Japan}
\affiliation{ESICMM, National Institute for Materials Science,
Tsukuba, Ibaraki 305-0047, Japan}
\affiliation{Center for Computational Sciences, University of Tsukuba,
Tsukuba, Ibaraki 305-8577, Japan}

\author{Takashi \surname{Miyake}}
\affiliation{CD-FMat, National Institute of Advanced Industrial Science
and Technology, Tsukuba, Ibaraki 305-8568, Japan}
\affiliation{ESICMM, National Institute for Materials Science,
Tsukuba, Ibaraki 305-0047, Japan}

\date{\today}
\begin{abstract}
We investigate the effects of various dopants ($M$ = K--Br) 
on the Curie temperature 
of the magnetic compound NdFe$_{12}$ through first-principles
calculations. 
Analysis by the Korringa--Kohn--Rostoker method with the coherent potential approximation reveals that 
doping the Fe sites with optimal concentrations of Ge and As is a promising strategy for increasing the Curie temperature.
To search over a wider space, we also perform
Bayesian optimization. 
Out of over 180,000 candidate compositions, 
co-doped systems with Co, Ge, and As are found to have the highest 
Curie temperatures.
\end{abstract}
\pacs{TBD}
\preprint{Ver.0.4.4}
\maketitle

\section{Introduction}
\label{S:Introduction}
Rare-earth--iron compounds are used in the highest-performance permanent magnets
currently available. 
The magnetic compound Nd$_2$Fe$_{14}$B is especially well known 
as the main phase of neodymium magnets,
which are the strongest magnets used industrially. 
In such magnets, rare-earth elements are the main source 
of the magnetic anisotropy and iron is the main source of the magnetic moment.

Compounds with the ThMn$_{12}$ structure are considered
promising because they can accommodate a larger
amount of Fe than Nd$_2$Fe$_{14}$B.
Hirayama et al. synthesized films of NdFe$_{12}$ and
reported that the nitrogenated film exhibited larger magnetization and 
higher Curie temperature than Nd$_2$Fe$_{14}$B.\cite{Hirayama15,Hirayama15b} 
However, NdFe$_{12}$(N) is not thermodynamically stable.

Doping of $R$Fe$_{12}$ has been investigated as a method 
for stabilizing the structure as a bulk material and 
enhancing the magnetic properties.
Optimization of the material properties by changing the composition of the system
is one of the central issues in the field.

Titanium energetically stabilizes the structure, and
an Fe-rich magnetic compound with a ThMn$_{12}$-type 
structure was first found as a Ti-doped system.\cite{Ohashi87,Ohashi88}
However, Ti also greatly reduces the magnetization of the system
owing to its antiferromagnetic coupling to the host Fe.\cite{Harashima16}
Cobalt is a typical enhancer of finite-temperature magnetism.
With respect to ThMn$_{12}$ compounds, 
Hirayama et al. reported
the synthesis of Co-doped Sm(Fe,Co)$_{12}$ films that
displayed excellent magnetic properties at room temperature
and a higher Curie temperature than the pristine system.\cite{Hirayama17}
First-principles calculations have
suggested that
Co not only improves the magnetic properties but also 
contributes to
the stability of the 
ThMn$_{12}$ structure.\cite{Harashima16}
We also discussed the enhancement of the Curie temperature, $T_\mathrm{C}$, 
and demonstrated that Cr is a better enhancer of this parameter
than Co in $R$Fe$_{12}$ ($R$=Y, Nd, Sm)
when the dopant concentration is low.\cite{Fukazawa18}
Using V, which is adjacent to Cr in the periodic table,
Sch\"{o}nh\"{o}bel et al.
synthesized SmFe$_{11}V$ and 
reported that its Curie temperature was 635 K,\cite{Schoenhoebel19}
which is significantly higher than the value of 555 K for 
SmFe$_{12}$.\cite{Hirayama17}

These works motivated us to explore a wider composition space for $T_\mathrm{C}$-enhancing dopants. 
We have recently developed a Bayesian optimization framework for such exploration
and demonstrated that it can greatly reduce the number of first-principles 
calculations required to identify the optimal system from a large candidate set.\cite{Fukazawa19c}
However, we considered only a few dopant elements in our previous study.
In this paper,
we examine
a series of dopants, namely, $M$ = K, Ca, Sc, Ti, V, Cr, Mn,
Co, Ni, Cu, Zn, Ga, Ge, As, Se, and Br,
as potential enhancers of the Curie temperature through first-principles calculations 
and also consider co-doping with some of these elements.

In Section~\ref{S:Methods}, we describe the details of the calculations.
As the first step, we performed first-principles calculations of NdFe$_{12-\delta}M_{\delta}$
for all of the dopants 
in the dilute limit of $M$ $(\delta \ll 1)$.
In Section~\ref{SS:Results_1}, we discuss how the dopants affect
the Curie temperature.
On the basis of the results, we selected six dopants
(V, Cr, Co, Ni, Ge, and As) with the potential to enhance 
the Curie temperature,
and we examine how a single dopant
changes the Curie temperature at a finite concentration
in Section~\ref{SS:Results_2}.
We also consider the effects of Ge and As as dopants
on the Curie temperature in terms of
hybridization between the Fe 3d and $M$ 4p orbitals.

In Section~\ref{SS:Results_3}, we consider the case of multiple dopants
and report on their advantages over a single dopant.
In the case of co-doping, the number of possible 
combinations becomes very large. 
To deal with this difficulty,
we applied 
the Bayesian optimization framework that we proposed previously.\cite{Fukazawa19c}
We demonstrate that co-doping with Co, Ge, and As has an advantage over
doping with Co alone.
Finally, we present our conclusions in Section \ref{S:Conclusion}.

\section{Methods}
\label{S:Methods}
We performed first-principles electronic structure calculations
based on density functional theory.\cite{Hohenberg64,Kohn65}
We used the Korringa--Kohn--Rostoker\cite{Korringa47, Kohn54} (KKR) Green's
function method to solve the Kohn--Sham equations
and the local density approximation for 
the exchange--correlation functional.
Although the spin--orbit coupling was not explicitly included
in the energy functional, the electronic configuration of 
the f electrons at the Nd site was assumed to obey 
Hund's rule. The f electrons were treated with the open-core
approximation and the self-interaction correction\cite{Perdew81}
was applied to the f states.
The randomness due to the occupation of dopants was treated within
the coherent potential approximation (CPA).\cite{Soven67,Soven70,Shiba71}

\begin{figure}[!t]
	\centering
	 \includegraphics[width=8cm]{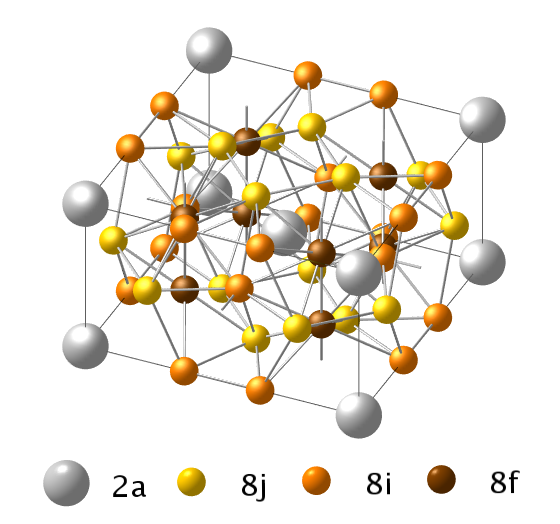}
	 \caption{\label{Fig:ThMn12} Structure of a
	ThMn$_{12}$-type crystal and its Wyckoff positions.}
\end{figure}

We assumed that NdFe$_{12}$ possesses the crystal structure of 
ThMn$_{12}$ [space group: I4/mmm (\#139)] (Fig.~\ref{Fig:ThMn12}),
and we adopted the lattice constant of NdFe$_{12}$
previously obtained \cite{Harashima15g}
for undoped and doped systems.

The Curie temperature was calculated from a classical 
Heisenberg model whose parameters were determined 
using Liechtenstein's method \cite{Liechtenstein87}
within the mean-field approximation.
Although this method overestimates the Curie temperature,
the relative changes among Fe-rich magnetic compounds are 
adequately described.\cite{Fukazawa19b}
Readers are referred to Ref.~\onlinecite{Fukazawa18}
for the details of the calculations.

We considered the doping of the Fe(8f), Fe(8i), and Fe(8j) sites
in NdFe$_{12}$. The magnetic moment of the dopant $M$
was assumed to be parallel or antiparallel to the total magnetic
moment of the system. Thus, we performed calculations with
the initial magnetic moment of the dopant
set parallel and antiparallel and carefully checked for 
the existence of metastable states.
The results for NdFe$_{12-\delta}M_{\delta}$ at an 
infinitesimal concentration 
(i.e., the dilute limit: $\delta \ll 1$) were obtained by fitting 
the data for $\delta$ = 0, 0.04, 0.08, 0.12, and 0.16
with polynomial curves.
The derivatives of the physical quantities are calculated
from the results.

To consider the case of co-doping with multiple elements,
we used a Bayesian framework for composition optimization
to identify the optimal system from a large search space.
The search framework was previously described in 
Ref.~\onlinecite{Fukazawa19c}.
For the Bayesian optimization,
we used the COMBO package, which can accommodate
a large number of candidates.\cite{Ueno16,COMBO}
The choice of system to explore next was conducted 
by Thompson sampling
after the initial 20 systems had been chosen at random.
The dimensionality of the random feature maps was set to 
2000.

\section{Results and Discussion}
\label{S:Results}
%
%
\subsection{NdFe$_{12-\delta}M_\delta\ (\delta \ll 1)$}
\label{SS:Results_1}
\begin{figure}[!t]
	\centering
	 \includegraphics[width=8cm]{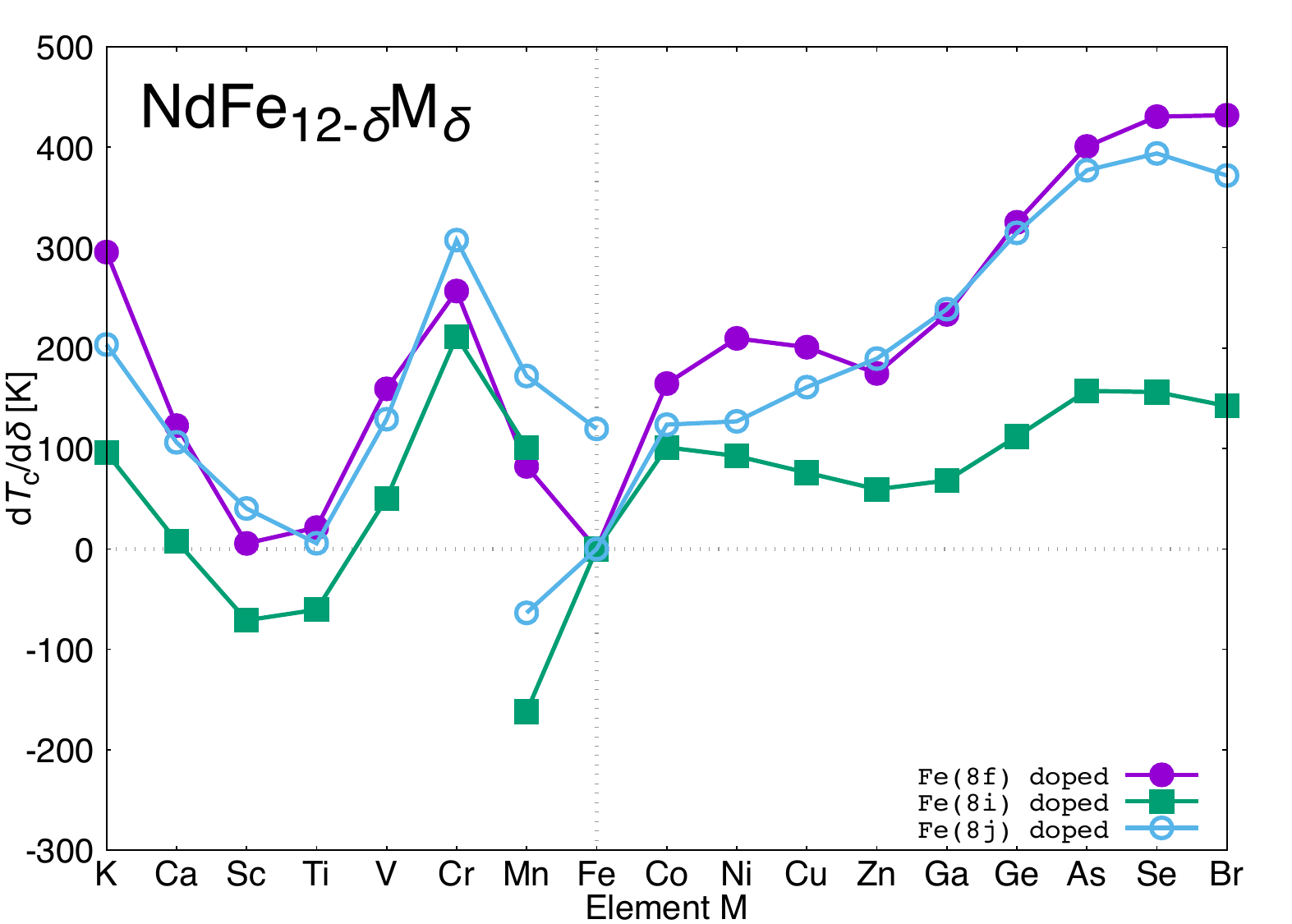}
	 \caption{\label{Fig:dTc_dc}
	 Derivative of the Curie temperature for
	 Nd(Fe$_{12-\delta}M_\delta$)
	 with respect to $\delta$
	 in the dilute limit, $\delta \rightarrow 0$.
	 These values are
	 the sum of the direct contribution shown in Fig.~\ref{Fig:dTc_dc_Direct}
	 and the indirect contribution shown in Fig.~\ref{Fig:dTc_dc_Indirect}. }
\end{figure}

In this subsection, we
present the results for NdFe$_{12-\delta}M_{\delta}$
in the dilute limit of $M$ ($\delta \ll 1$).
Figure~\ref{Fig:dTc_dc} shows the derivative of
$T_{\mathrm C}$ with respect to concentration,
$\left.\frac{dT_{\mathrm C}}{d\delta}\right|_{\delta=0}$.
In a previous paper,\cite{Fukazawa18}
we demonstrated the potential of Cr 
in enhancing the Curie temperature of
$R$Fe$_{12}$ ($R$=Y, Nd, Sm) 
more efficiently than Co for low dopant concentrations.
The curve of the derivative shown in the figure exhibits 
a peak at $M$=Cr.
We also see that there are significant increases 
for both $M$=K and $M$=Ge--Br.

To analyze the origin of the enhancement,
we performed direct--indirect decomposition (DID), 
which we previously proposed.\cite{Fukazawa18}
The Curie temperature can be calculated from the
intersite couplings $J_{ij}$ and concentration $\delta$
within the mean-field approximation.
Because $J_{ij}$ is also a function of $\delta$, we see that
the change of $T_\mathrm{C}$ with respect to concentration can be expressed as
\begin{equation}
	\left.
	\frac{
		dT_\mathrm{C}[\{J_{ij}(\delta)\}, \delta]
	}
	{
		d\delta
	}
	\right|_{\delta=0}
	=
	\left.
	\frac{
		\partial T_\mathrm{C}
	}
	{
		\partial \delta
	}
	\right|_{\{J_{ij}(0)\}}
	+ 
	\sum_{ij}
	\left.
	\frac{
		\partial T_\mathrm{C}
	}
	{
		\partial J_{ij}
	}
	\frac{
		dJ_{ij}
	}
	{
		d\delta
	}
	\right|_{\delta=0}.
\end{equation}
The first and second terms are referred to as the direct and indirect parts, respectively.

\begin{figure}[!t]
	\centering
	 \includegraphics[width=8cm]{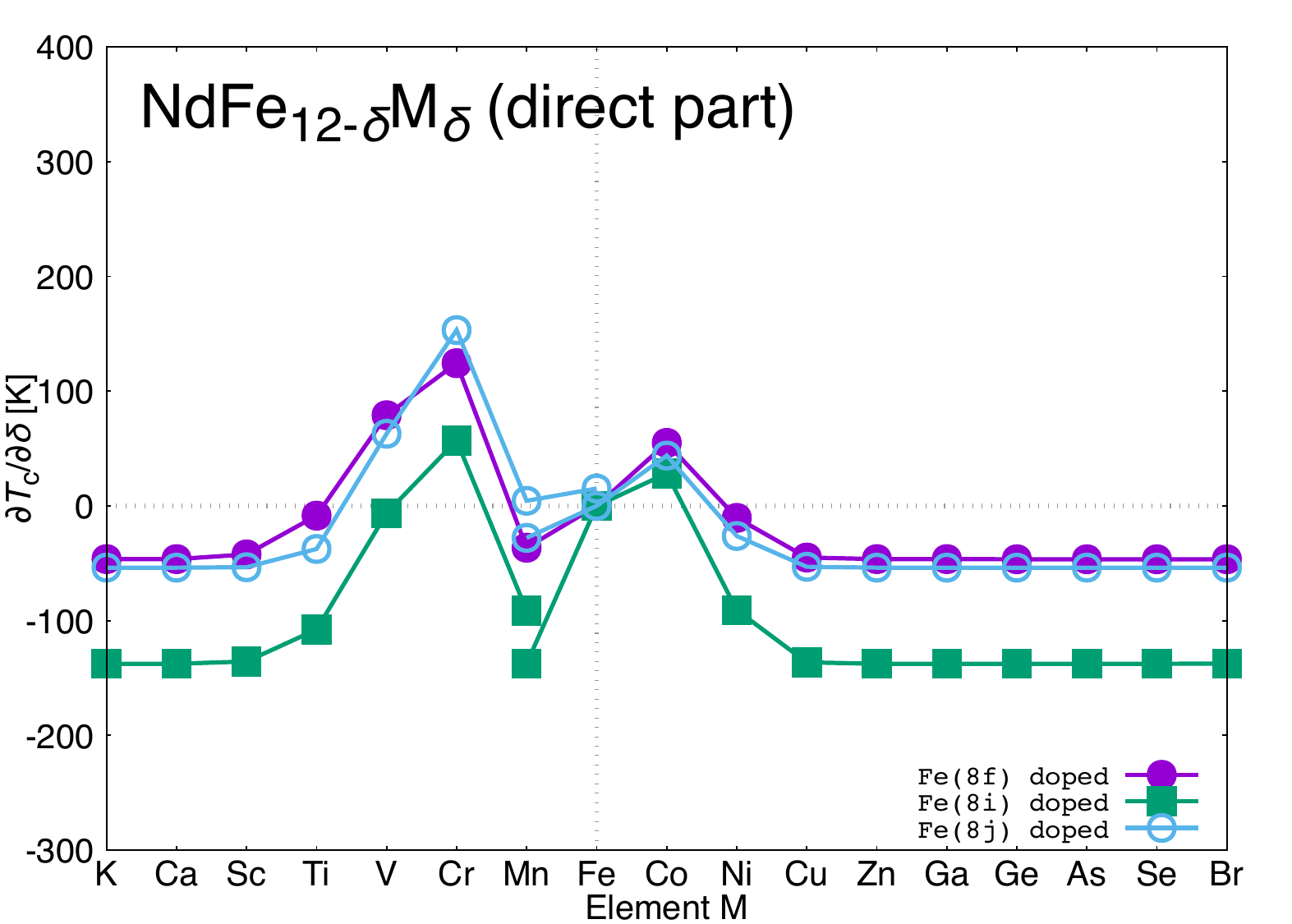}
	 \caption{\label{Fig:dTc_dc_Direct}
	 Direct contribution to 
	 the derivative of the Curie temperature for
	 Nd(Fe$_{12-\delta}M_\delta$)
	 in the dilute limit.}
\end{figure}

Figure~\ref{Fig:dTc_dc_Direct} shows
the direct contribution to the derivative of $T_\mathrm{C}$, which originates from
the difference in the magnetic couplings
between the replacing Fe--$M$ couplings and
the replaced Fe--Fe couplings.
The direct contribution is largely positive for $M$ = Co, Cr, and V,
whereas it is negative for most of the remaining cases.

\begin{figure}[!t]
	\centering
	 \includegraphics[width=8cm]{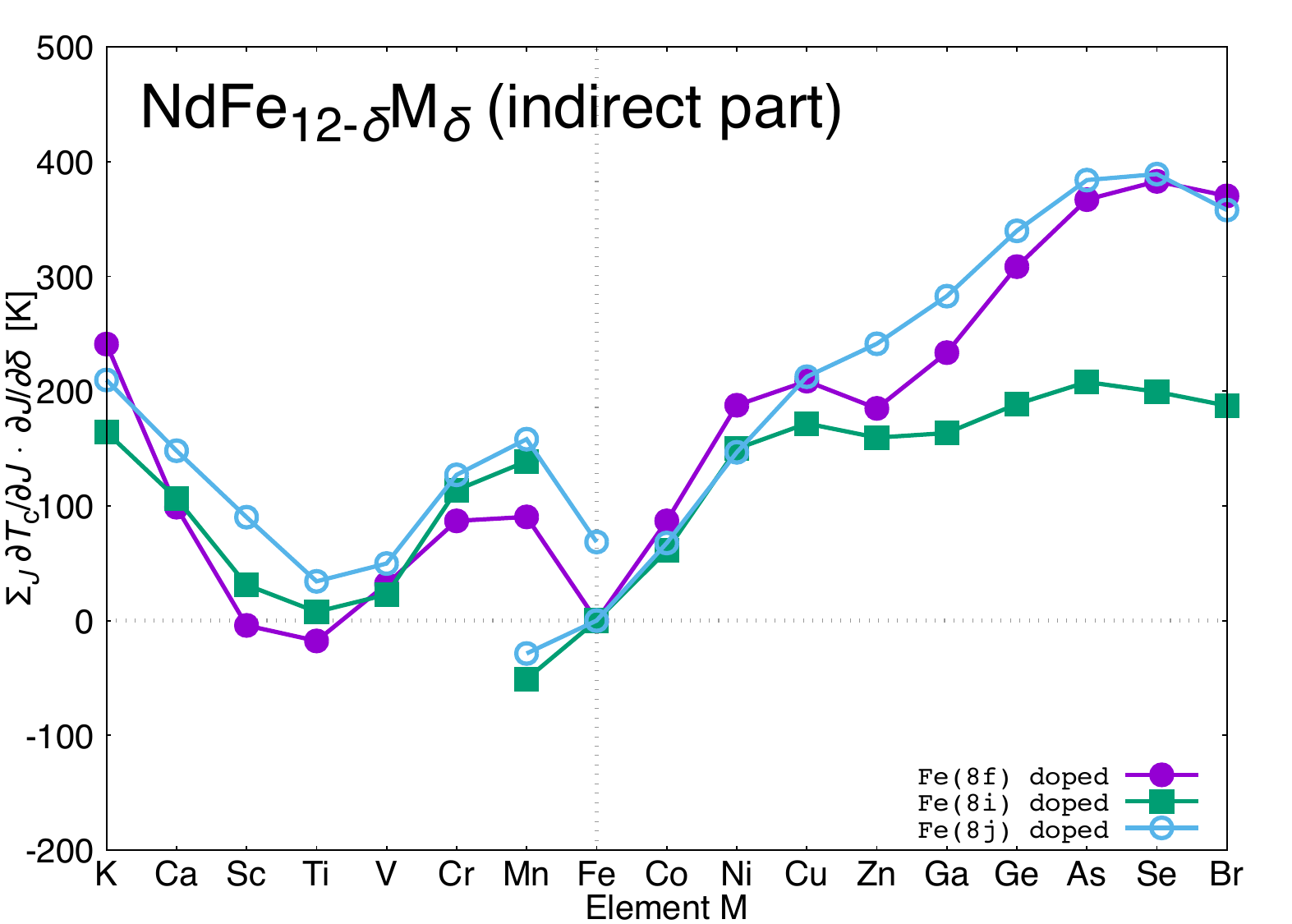}
	 \caption{\label{Fig:dTc_dc_Indirect}
	 Indirect contribution to 
	 the derivative of the Curie temperature for
	 Nd(Fe$_{12-\delta}M_\delta$)
	 in the dilute limit.}
\end{figure}

Figure~\ref{Fig:dTc_dc_Indirect} shows the indirect contribution obtained from
the DID, which originates from the enhancement of 
the magnetic Fe--Fe couplings due to the introduction of $M$.
It is noteworthy that 
the significant enhancement observed for $M$=K and Ge--Br in Fig.~\ref{Fig:dTc_dc}
can be attributed solely to the indirect contribution.
We discuss the atomic-scale origin of this enhancement 
in terms of hybridization between the Fe 3d
and $M$ 4p orbitals in the next subsection.

\begin{figure}[!t]
	\centering
	 \includegraphics[width=8cm]{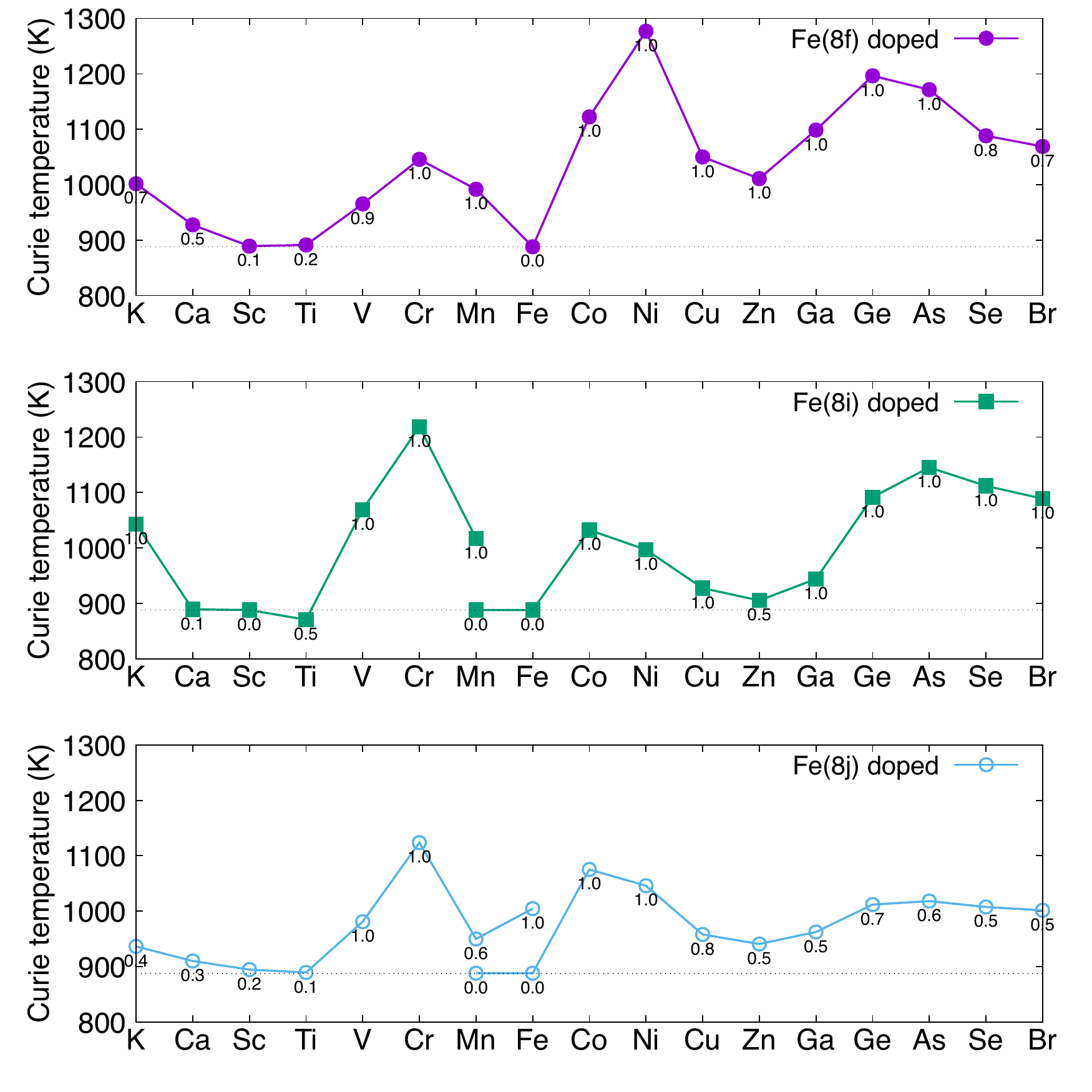}
	 \caption{\label{Fig:Tc_quad_peak}
	Estimated maximum Curie temperatures
	for Nd(Fe$_{12-\delta}$M$_\delta$)
	in the range of $0 \leq \delta \leq 1$
	obtained from the quadratic model.
	The numbers below each data point 
	indicate the corresponding values of $\delta$.}
\end{figure}

To roughly estimate the Curie temperature 
for finite dopant concentrations, 
we constructed
a quadratic model of $T_{\mathrm C}$
as a function of the concentration
through estimation of the first and second derivatives by data fitting:
\begin{equation}
    T_{\mathrm C}(\delta)
    =
    T_{\mathrm C}(0)
    +
    \frac{dT_{\mathrm C}}{d\delta}(0)\,
    \delta
    +
    \frac{1}{2}
    \frac{d^2T_{\mathrm C}}{d\delta^2}(0)\,
    \delta^2.
    \label{Eq:Tc_Model}
\end{equation}
This model is valid for low dopant concentrations.
Figure~\ref{Fig:Tc_quad_peak} presents the highest values of $T_{\mathrm C}$
for Nd(Fe$_{12-\delta}$M$_\delta$) in the range of 
$0 \leq \delta \leq 1$ as a function of $M$.
The numbers below each data point indicate
the concentrations
that afforded the highest values of $T_{\mathrm C}$.
From the figure, we can expect a large enhancement of 
the Curie temperature by 
doping with $M$ = V, Cr, Co, Ni, Ge, or As alone.

\subsection{Finite amount of $M$}
\label{SS:Results_2}
In this subsection, we investigate the
Curie temperature of
Nd(Fe$_{12-x}$M$_x$) for a finite concentration $x$
(we let $x$ denote the concentration to indicate that
it is finite).
First, we consider the results for $M$ = V, Cr, Co, Ni, Ge, and As
(which we selected on the basis of Fig.~\ref{Fig:Tc_quad_peak})
to examine the validity of the rough estimation described above.
To obtain the data, we performed calculations for
finite concentrations of $x=0, 0.25, ..., 2$.

\begin{figure*}[!t]
    \centering
    \includegraphics[width=35mm]{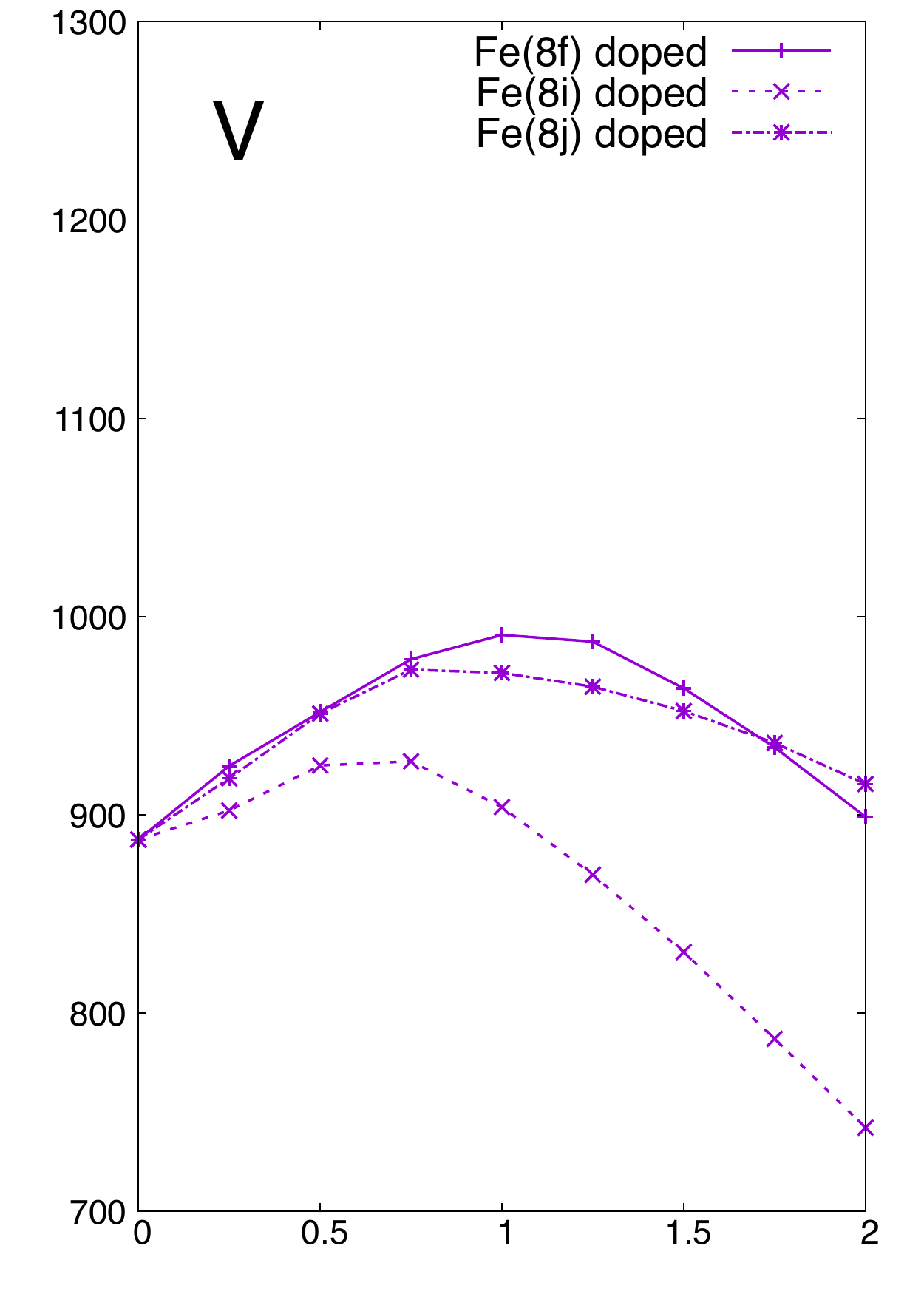}
	\includegraphics[width=35mm]{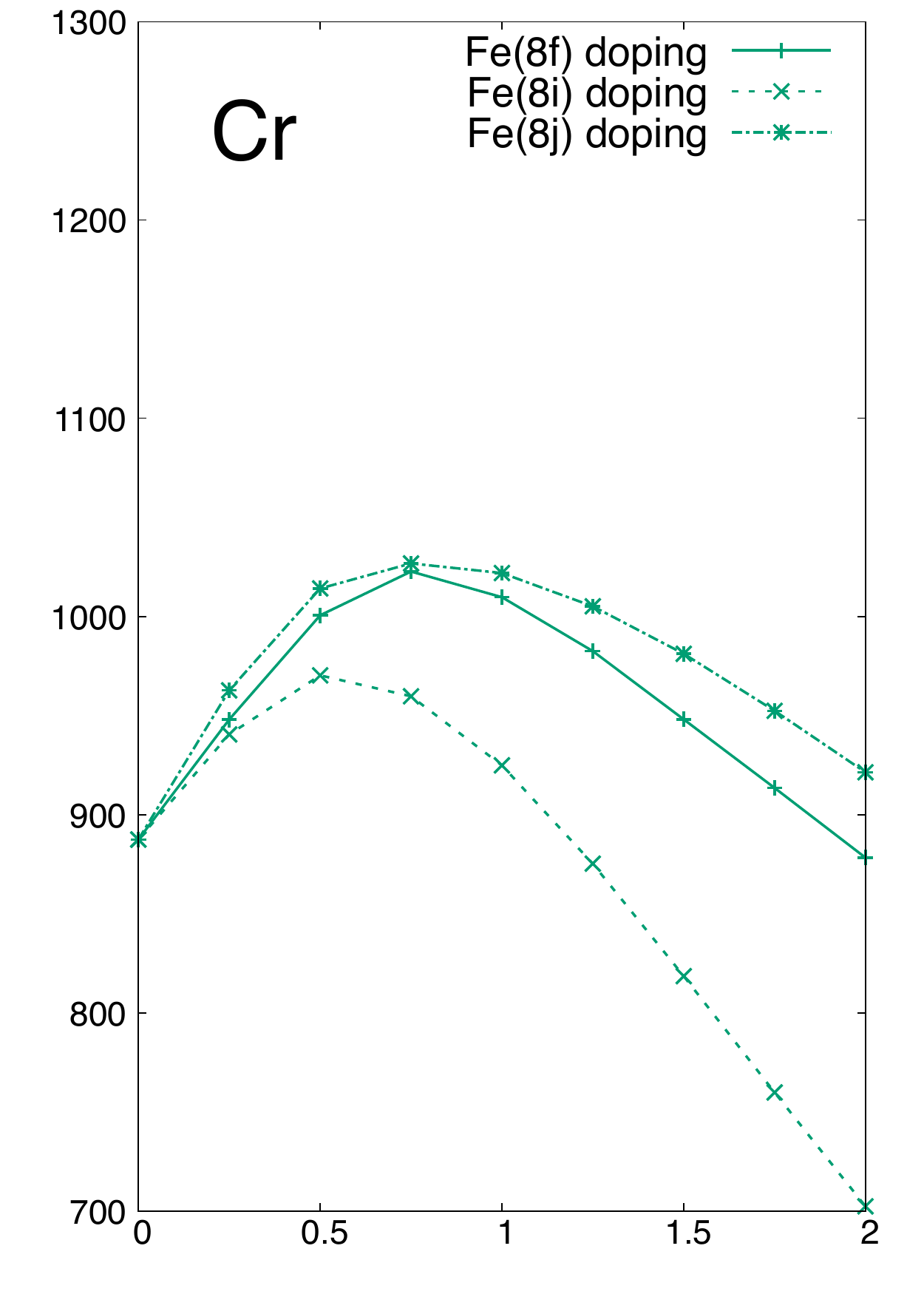}
    \includegraphics[width=35mm]{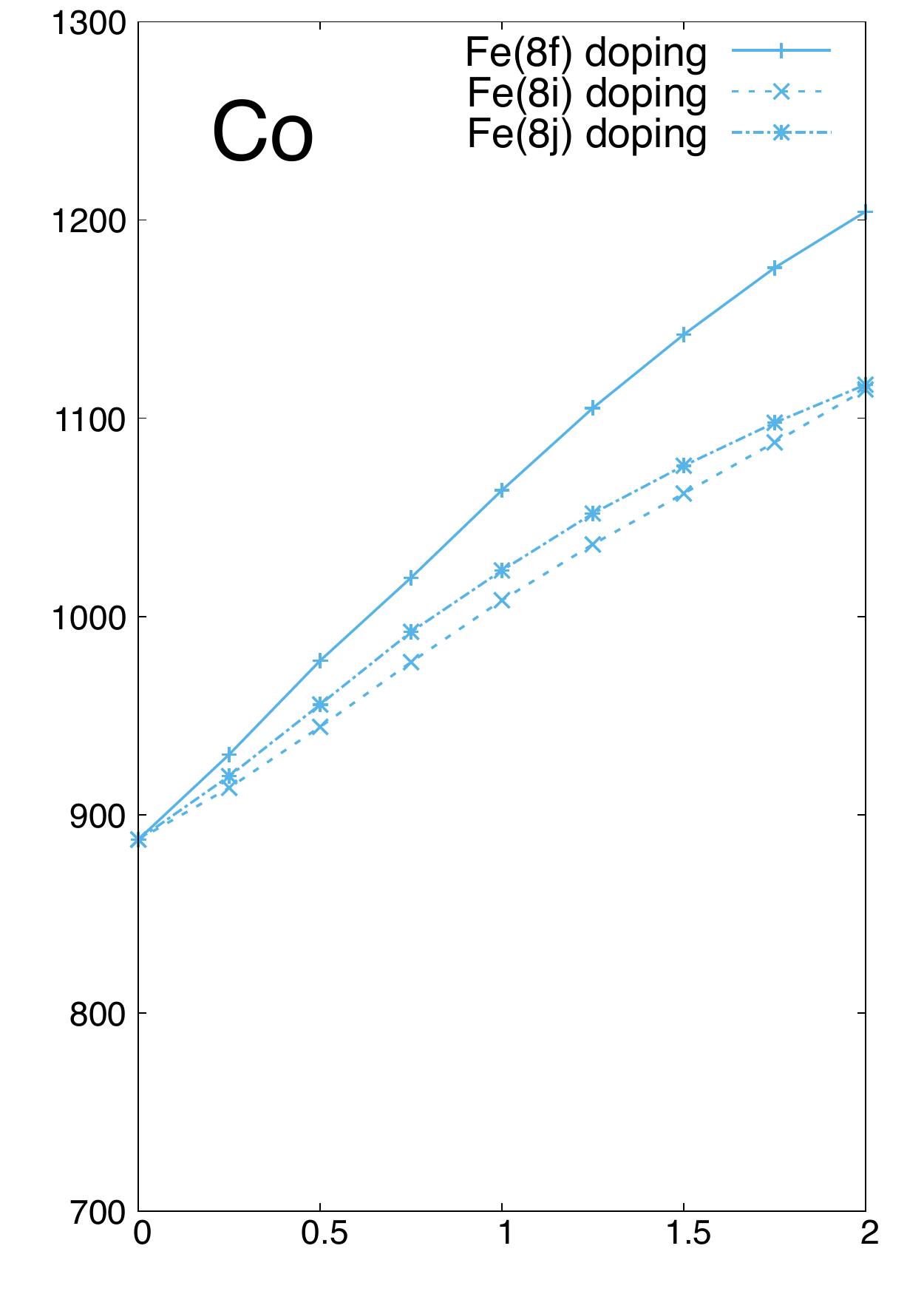}
    \\
	\includegraphics[width=35mm]{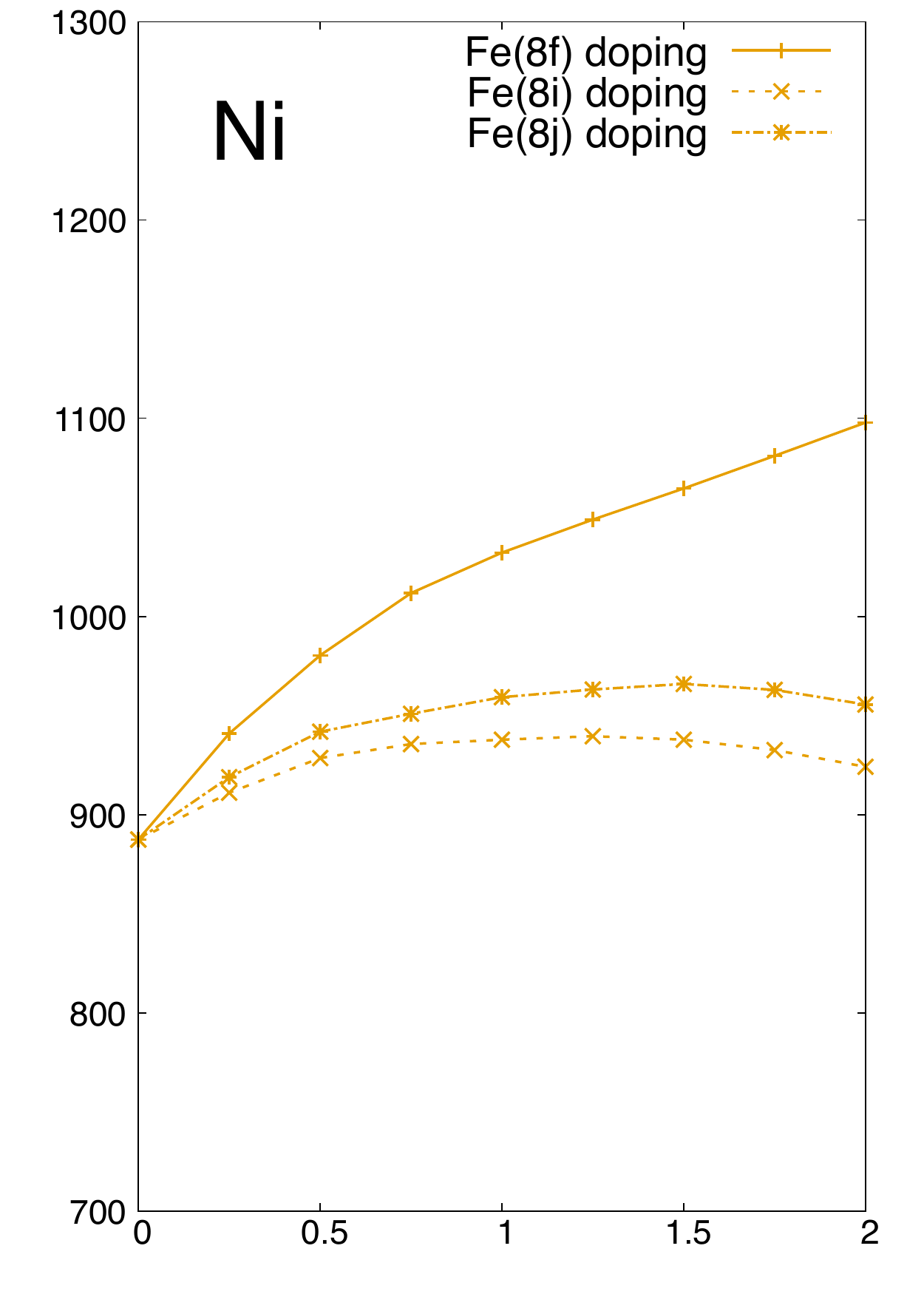}
	\includegraphics[width=35mm]{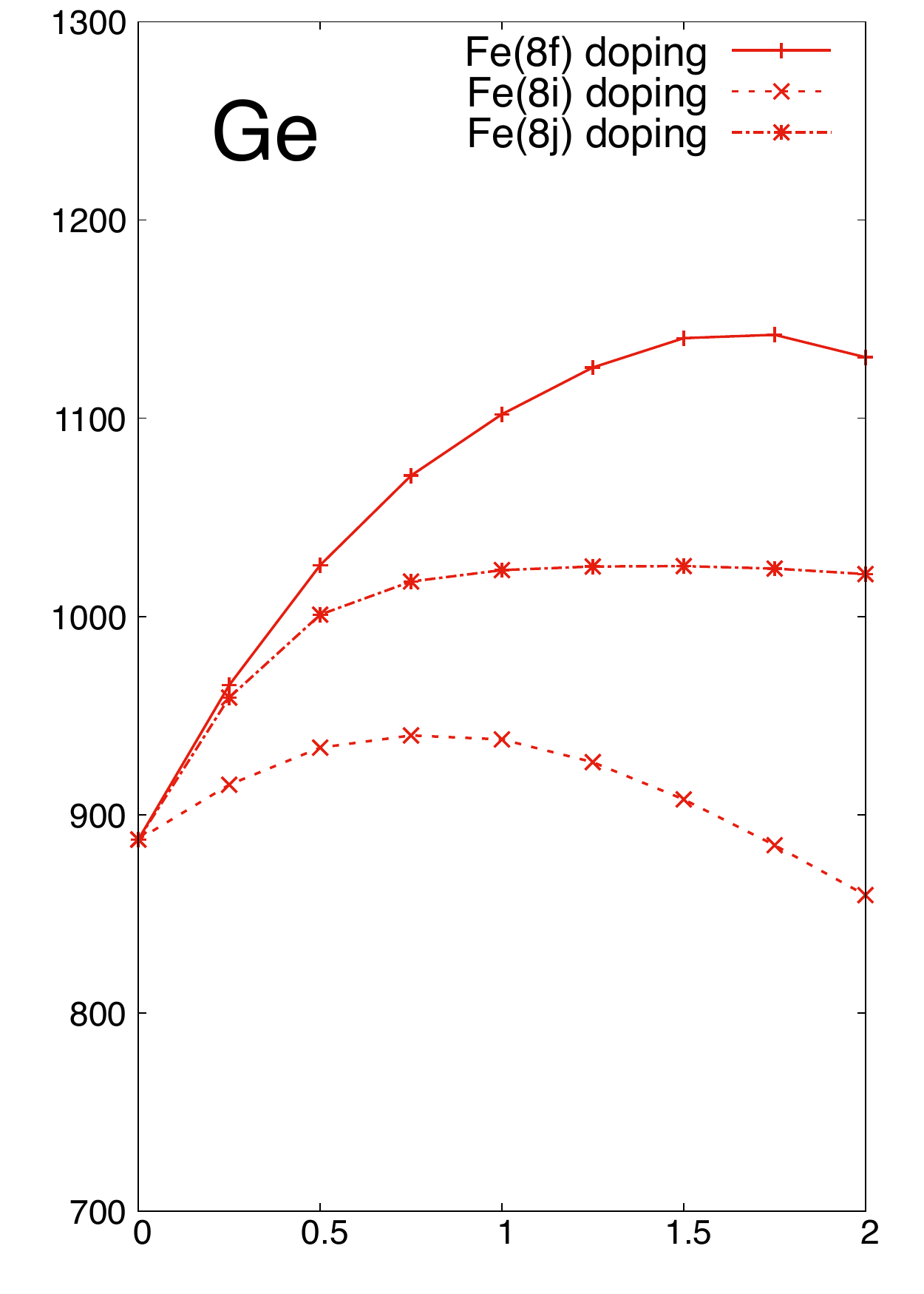}
	\includegraphics[width=35mm]{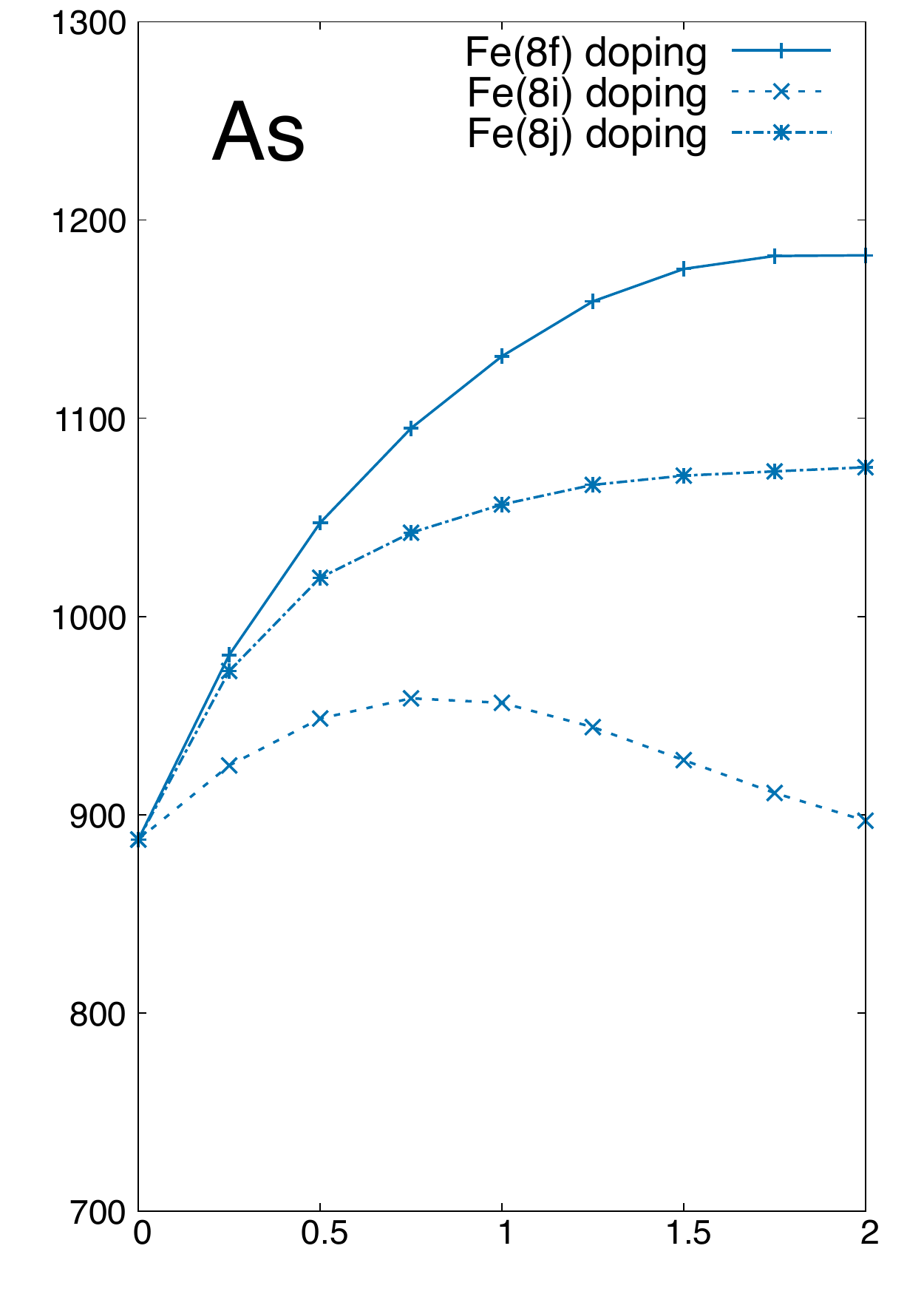}
	 \caption{\label{Fig:Tc_finite_conc}
	 Curie temperatures for Nd(Fe$_{12-x}$M$_x$) (M=V, Cr, Co, Ni, Ge, or As)
	 with finite concentrations $x$ in the range of $0 \leq x \leq 2$.}
\end{figure*}

Figure~\ref{Fig:Tc_finite_conc} shows the calculated values
of the Curie temperature.
It follows from this figure that 
the maximum $T_\mathrm{C}$ was overestimated by the results shown 
in Fig.~\ref{Fig:Tc_quad_peak}; however, 
the relative changes among the dopants are described well.
For V and Cr, $T_\mathrm{C}$ begins to decrease
around $x=1$. In the case of Co, the maximum $T_\mathrm{C}$ occurs outside the range of the figure. Ni is less favorable for maintaining $T_\mathrm{C}$.
Ge and As display similar curves, where the maximum values of
$T_\mathrm{C}$ occur at a higher concentration than those observed for V and Cr.

As we have previously reported,\cite{Fukazawa18}
Cr is a better enhancer
of $T_\mathrm{C}$ than Co for low dopant concentrations.
This is attributable to the strong antiferromagnetic Fe--Cr coupling,
which leads to the direct contribution shown
in Fig.~\ref{Fig:dTc_dc_Direct}.
Vanadium plays a similar role owing to
the strong Fe--V coupling, although it is less effective
than Cr. In the case of Cr or V doping, the second-order effects of $x$ 
decrease the Curie temperature and cause the curve to first 
increase and then decrease.

For $M$=Co or Ni, the dopants largely reinforce
the magnetic Fe--Fe couplings,\cite{Fukazawa18}
which is 
the indirect contribution.
The second-order effects are weak 
because the moment of $M$ is parallel to the host and 
the $M$--$M$ coupling is ferromagnetic.
Therefore, Co and Ni can serve as enhancers of $T_\mathrm{C}$
over a wider range of concentrations.

In the cases of $M$=Ge and As, the enhancement of
the Curie temperature can be ascribed to
the indirect contribution as shown in 
Fig.~\ref{Fig:dTc_dc_Indirect}.
This effect can be explained in terms of
hybridization between the Fe 3d
and $M$ 4p states.
It is also noteworthy that 
the maxima for $M$=Ge and As
occur at higher concentrations than for V and Cr and lower concentrations than for Co and Ni
because Ge and As are non-magnetic dopants.

\begin{figure}[!t]
	\centering
	\includegraphics[width=65mm]{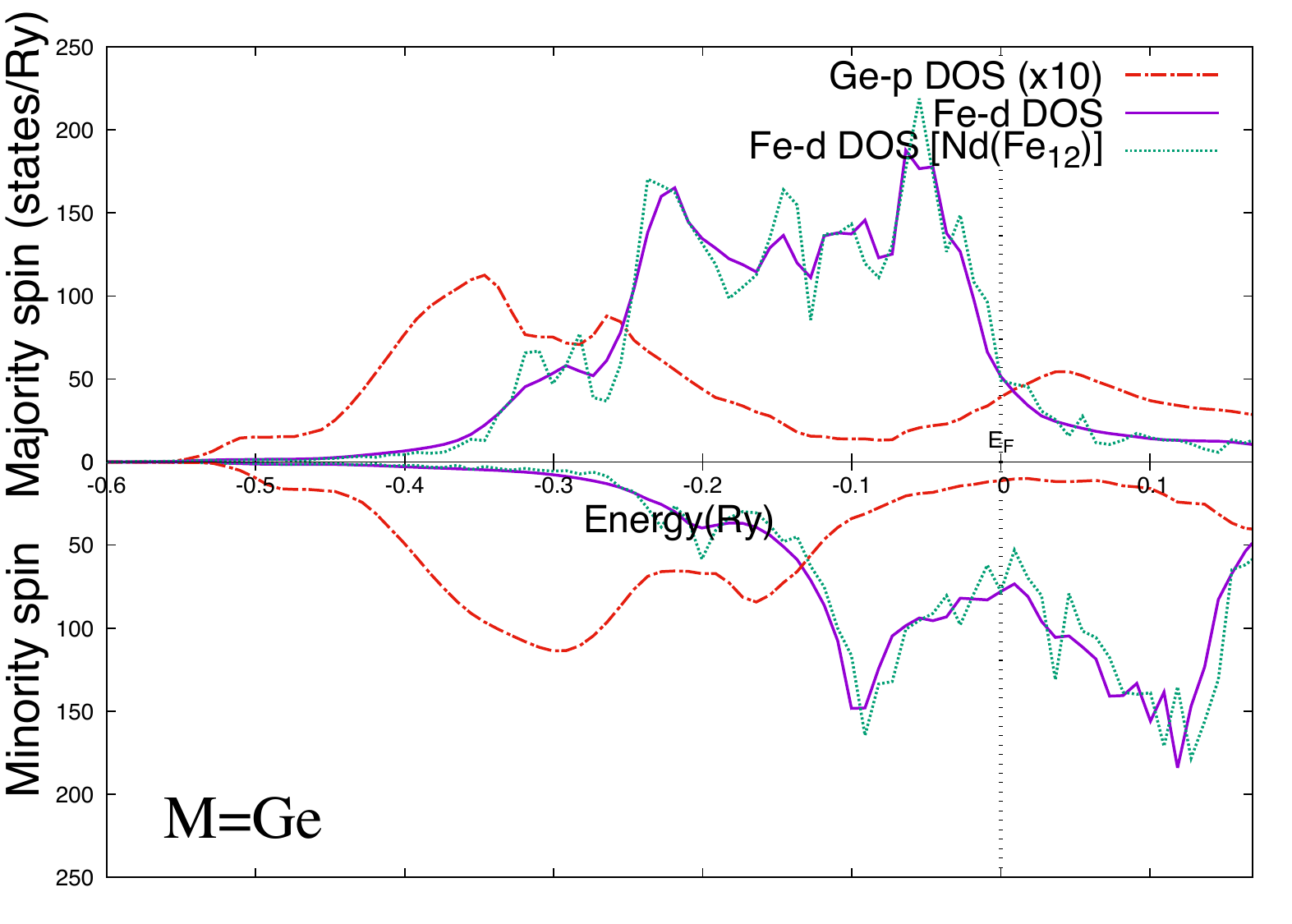}
	\includegraphics[width=65mm]{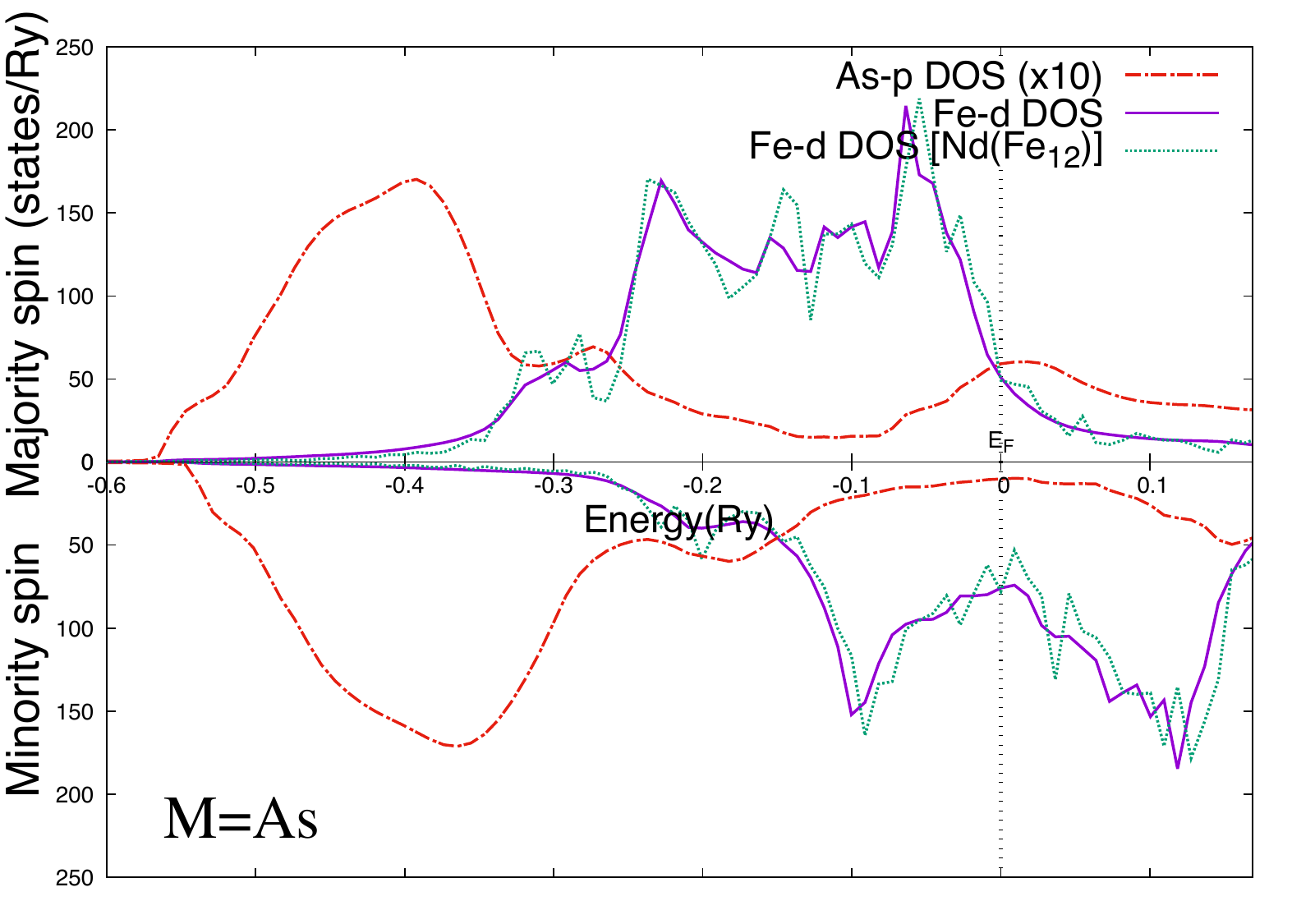}
	\includegraphics[width=65mm]{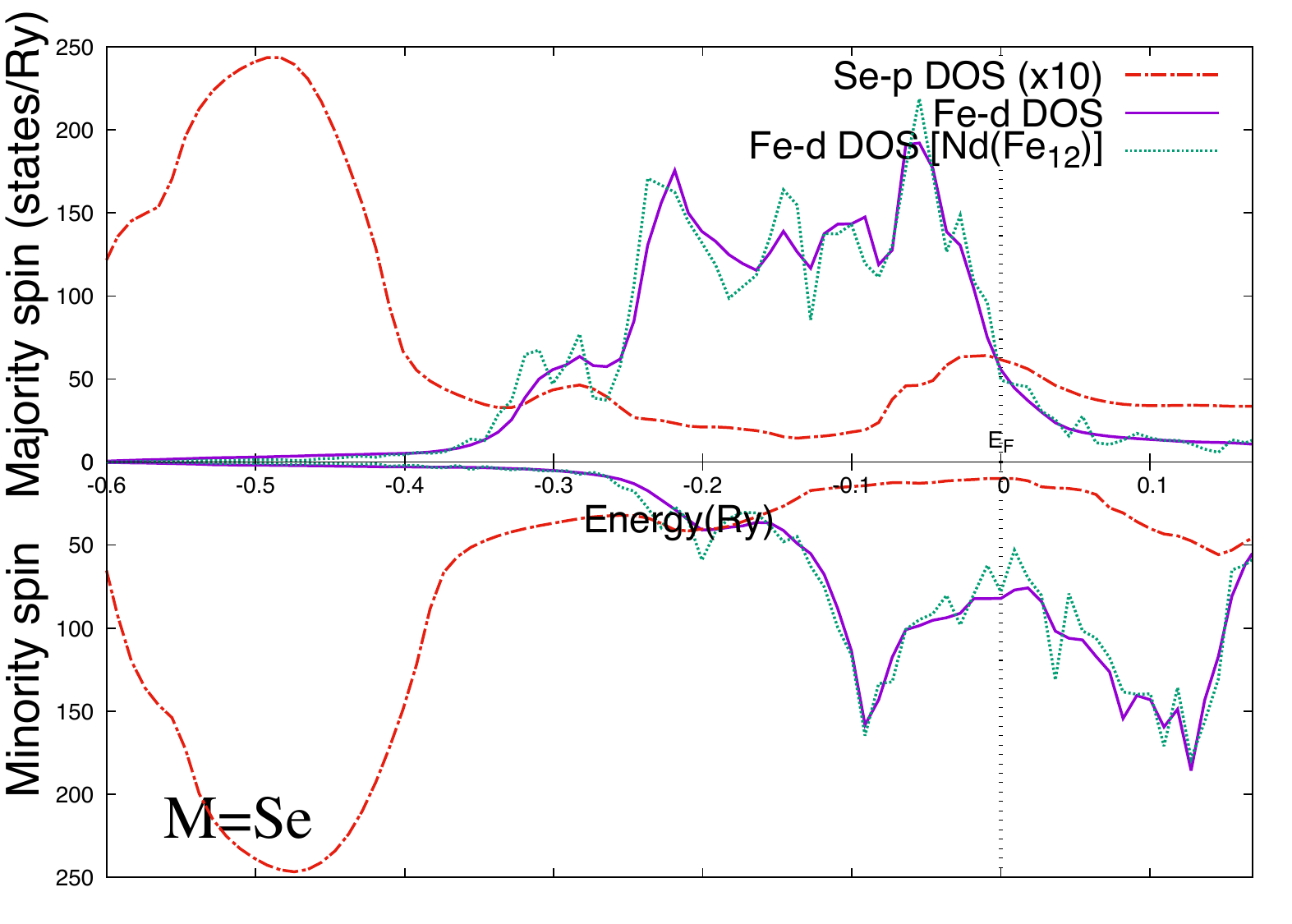}
	\includegraphics[width=65mm]{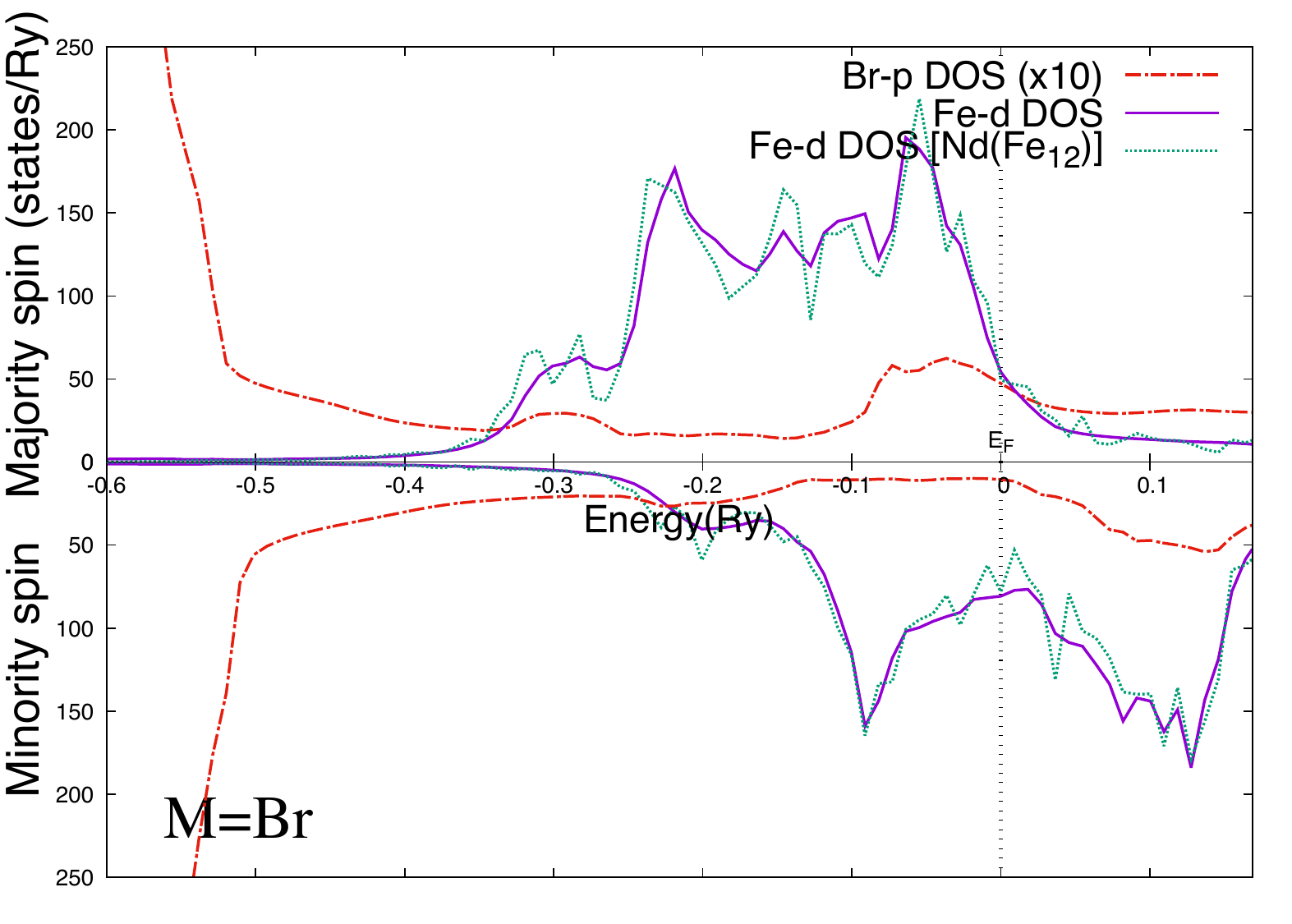}
	 \caption{\label{Fig:DOSCOM_finite_conc}
	Partial DOSs for the Fe d orbitals and M p orbitals (multiplied by ten)
	of the 8f-doped Nd(Fe$_{11}$M) systems
	(M=Ge, As, Se, or Br). }
\end{figure}

To examine the chemical trends, 
we performed calculations for 8f-doped Nd(Fe$_{11}M$) 
with $M$=Ge, As, Se, and Br (atomic numbers 32--35).
Figure~\ref{Fig:DOSCOM_finite_conc} shows 
the partial densities of states (DOSs).
In the case of $M$=Ge, we can see from the DOS for the $M$ p orbitals that a large part of
the antibonding states remains unoccupied.
This situation is comparable to {\it cobaltization}, in which
the DOS of Fe is deformed to become similar to that of Co by hybridization 
with unoccupied states
at a neighboring site.\cite{Kanamori90,Ogura11,Harashima15e}
This is considered to reinforce the magnetism by 
strengthening the magnetic coupling between the Co-like Fe and
the surrounding Fe.

\begin{figure}[!t]
	\centering
	\includegraphics[width=7.5cm]{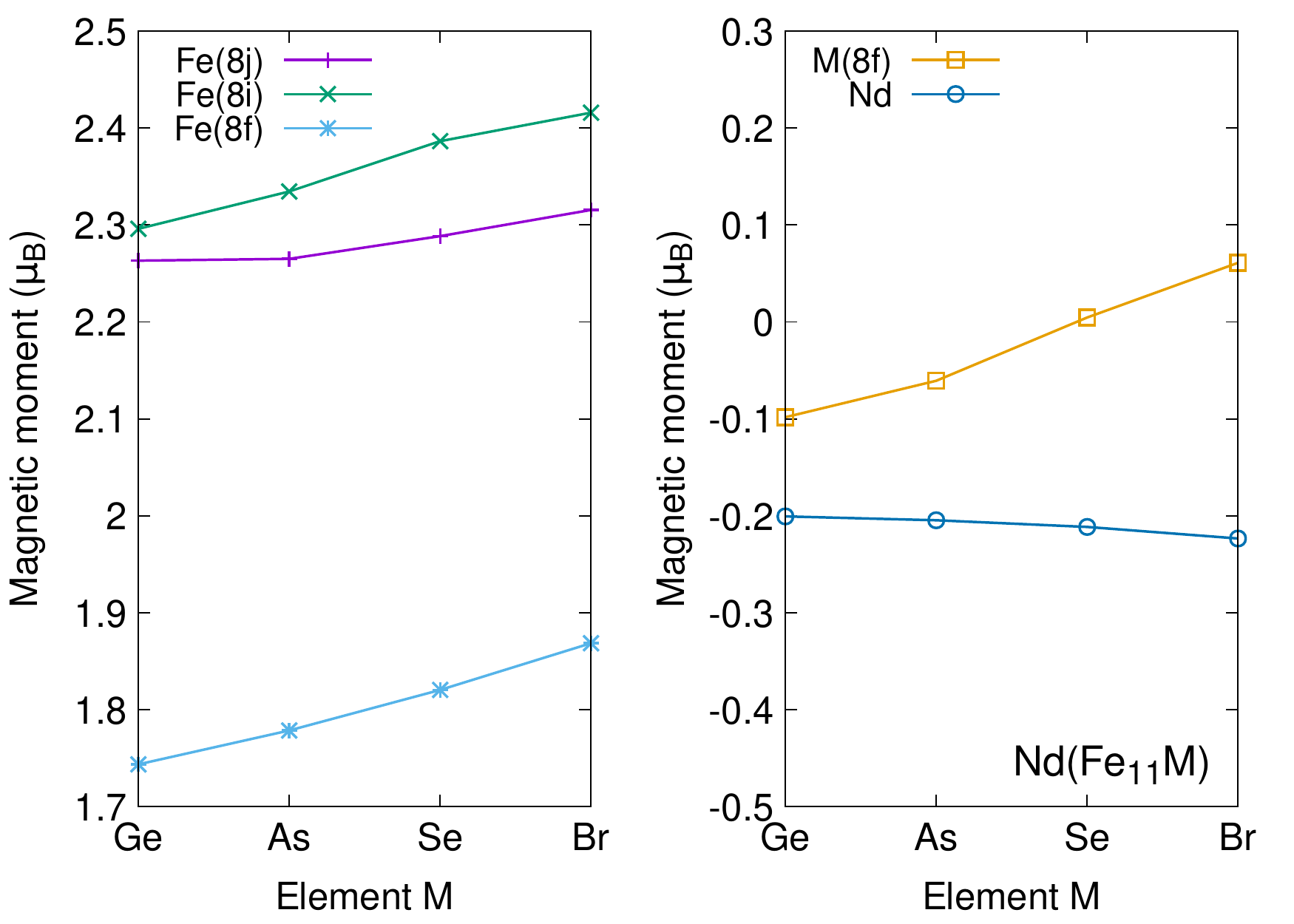}
	\caption{\label{Fig:0481_0485_local_moment}
	Local magnetic moments in 8f-doped Nd(Fe$_{11}$M) (M=Ge, As, Se, or Br).}
\end{figure}

\begin{figure}[!t]
	\centering
	\includegraphics[width=8.0cm]{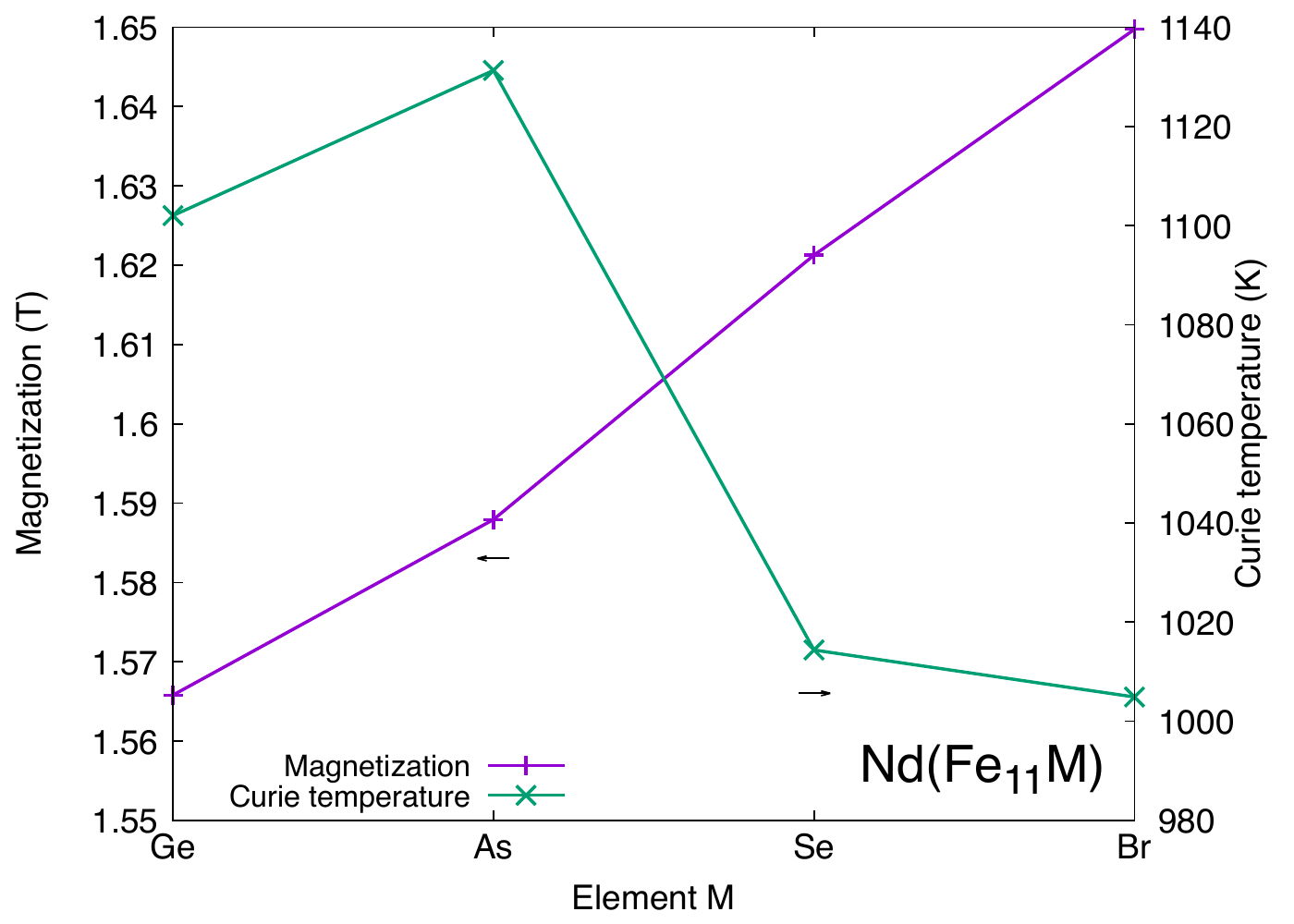}
	\caption{\label{Fig:0481_0485_tesla_Tc}
	Magnetization and Curie temperature for 
	8f-doped Nd(Fe$_{11}$M) (M=Ge, As, Se, or Br). }
\end{figure}

Upon increasing the atomic number of the dopant from $M$=Ge,
the potential becomes deeper and more majority-spin channels 
are occupied (Fig.~\ref{Fig:DOSCOM_finite_conc}),
while the minority-spin channels remain unoccupied.
This enhances the local moment of the Fe sites (Fig.~\ref{Fig:0481_0485_local_moment})
and makes the spin-rotational perturbation considered in
Liechtenstein's formula larger,
which increases the estimated values of the
intersite magnetic couplings and the Curie temperature.
However, 
the {\it cobaltization} is simultaneously weakened by the partial occupation 
of the antibonding states, which decreases the Curie temperature.
The crossover of these two effects
is responsible for the peak in $T_\mathrm{C}$ for $M$=As
observed in Fig.~\ref{Fig:0481_0485_tesla_Tc},
although the local moments of Fe monotonically increase with increasing atomic number  (Fig.~\ref{Fig:0481_0485_local_moment}).

\subsection{Systems with multiple dopants}
\label{SS:Results_3}
Finally, we consider the co-doping of NdFe$_{12}$ to enhance
the Curie temperature.
It should be readily apparent that 
the investigation of all possible combinations 
of $M$=K--Br would be impractical.
We avoided this problem by screening the dopants on the basis of the 
results shown in Fig.~\ref{Fig:Tc_quad_peak}; 
we hereinafter consider doping with Co, Ni, Ge, and As for the 8f site,
Cr and Co for the 8i site, and Cr, Co, and Ni for the 8j site.

We also focus on the regime of low dopant concentrations
because Fe-rich compounds are favorable in terms of magnetization,
and several of the dopants were expected to afford $T_\mathrm{C}$ maxima at low 
concentrations in the range of $x \leq 2$ (Fig.~\ref{Fig:Tc_finite_conc}). 
We prepared two lists of candidates, lists (A) and (B),
with different upper limits of dopant concentration.
In list (A), the amount of each dopant per formula was varied from 
$0$ to $1$ in intervals of $0.1$ with the constraint
that the total amount of dopants was $x \leq 1$.
In list (B), the amount of each dopant was varied from 
$0$ to $2$ in intervals of $0.2$ with the constraint
of $x \leq 2$.
Each list consisted of 92,378 systems and 
up to nine site--dopant combinations per system.
The total number of unique items is 182754 (2002 duplicates).

Even with this screening, considerable time and resources
would be required to perform first-principles calculations for all of the candidates.
To overcome this problem, we applied 
our efficient framework 
for optimization of the chemical composition
based on Bayesian optimization.\cite{Fukazawa19c}
With this framework, it is possible to identify high-performance materials
from a candidate list
with a small number of data acquisition processes
by alternately performing data acquisition and stochastic modeling.

\begin{figure}[!t]
	\centering
	\includegraphics[width=7.5cm]{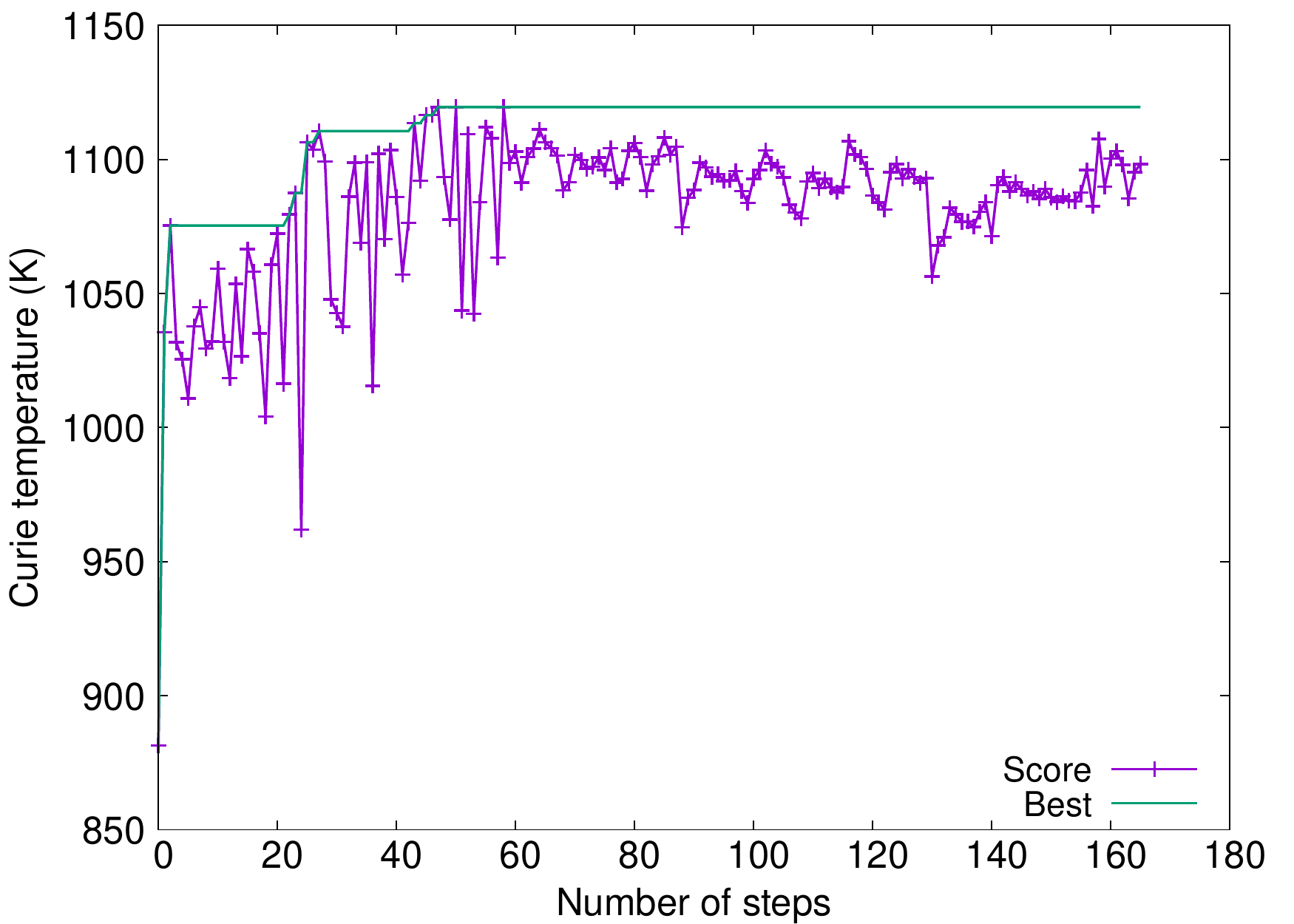}
	\includegraphics[width=7.5cm]{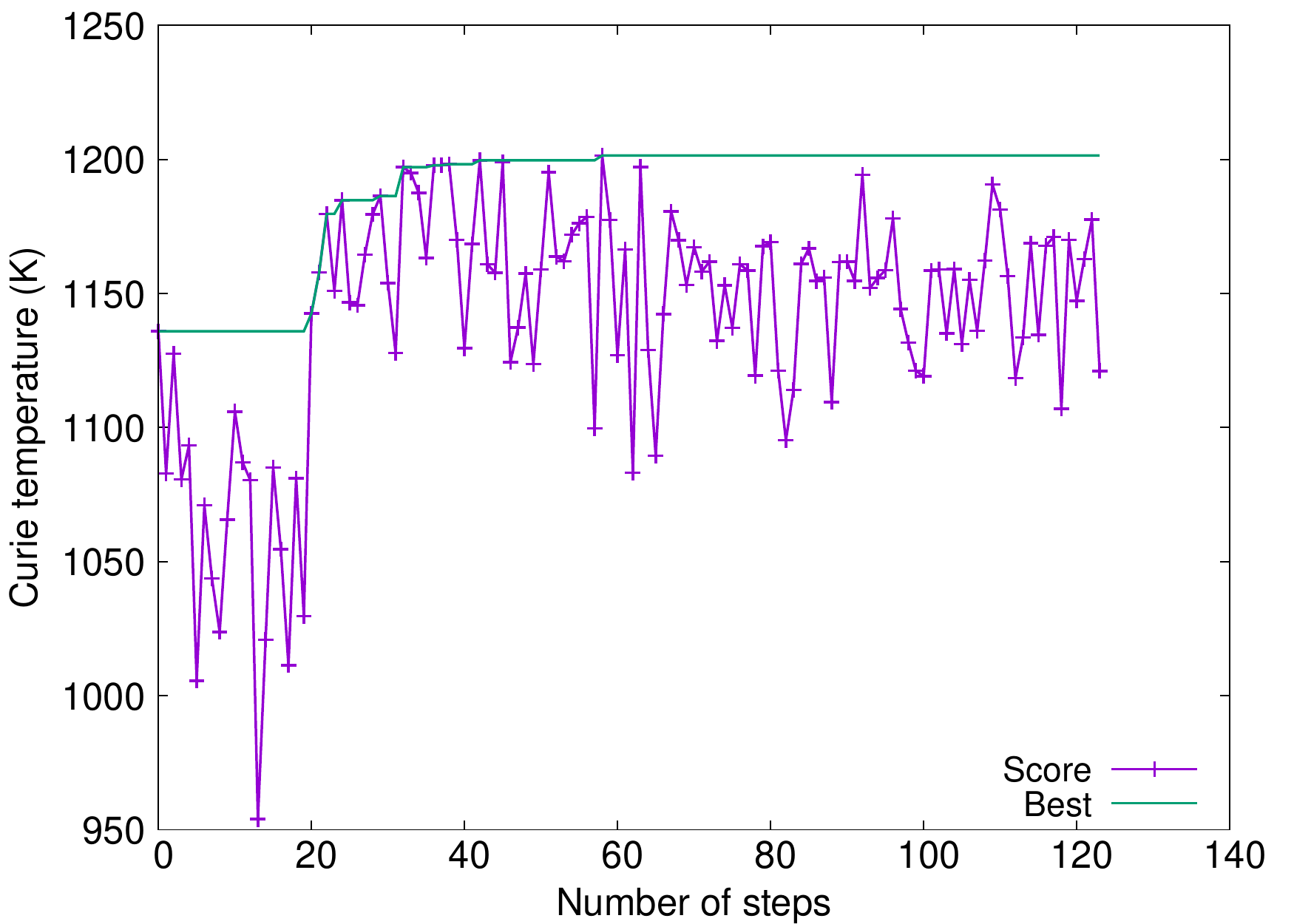}
	\caption{\label{Fig:nu_score}
	Values of the Curie temperature at each optimization step
	for list (A) (top) and list (B) (bottom). }
\end{figure}

Figure~\ref{Fig:nu_score} shows the obtained Curie temperature 
versus the number of data acquisition steps.
In these plots, the score ($T_\mathrm{C}$) at each step and 
cumulative best score are shown for lists (A) and (B).
In both cases, 
the best system observed during the run was found within the first 60 steps.
After this point, the score oscillated between lower values, 
indicating that there remained few or no better systems.

The ten best identified systems are shown in 
Table~\ref{Tbl:top10_opt1} for list (A)
and Table~\ref{Tbl:top10_opt2} for list (B).
\begin{table}
	\centering
	\caption{\label{Tbl:top10_opt1}
	Ten best systems found in the optimization using list (A).}
	\begin{tabular}{lcc}
		\hline
		Formula & $T_\mathrm{C}$ (K) & $\mu_0 M$ (T)\\
		\hline
		NdFe$_{11}$As & 1119 & 1.58\\
		NdFe$_{11}$As$_{0.9}$Ge$_{0.1}$ & 1117 & 1.58\\
		NdFe$_{11}$As$_{0.8}$Ge$_{0.2}$ & 1114 & 1.58\\
		NdFe$_{11}$As$_{0.9}$Co(8j)$_{0.1}$ & 1112 & 1.60\\
		NdFe$_{11}$As$_{0.8}$Co(8f)$_{0.2}$ & 1111 & 1.61\\
		NdFe$_{11}$As$_{0.7}$Ge$_{0.3}$ & 1111 & 1.58\\
		NdFe$_{11}$As$_{0.7}$Ge$_{0.2}$Co(8f)$_{0.1}$ & 1109 & 1.60\\
		NdFe$_{11}$As$_{0.7}$Ge$_{0.1}$Co(8f)$_{0.2}$ & 1108 & 1.61\\
		NdFe$_{11}$As$_{0.8}$Ge$_{0.1}$Co(8i)$_{0.1}$ & 1108 & 1.59\\
		NdFe$_{11}$As$_{0.6}$Ge$_{0.4}$ & 1108 & 1.58\\
		\hline
		NdFe$_{12}$ & 881 & 1.73 \\
		\hline
	\end{tabular}
\end{table}
\begin{table}
	\centering
	\caption{\label{Tbl:top10_opt2}Ten best systems found in the optimization using list (B).}
	\begin{tabular}{lcc}
		\hline
		Formula & $T_\mathrm{C}$ (K) & $\mu_0 M$ (T)\\
		\hline
		NdFe$_{10}$Co(8f)$_{1.8}$As$_{0.2}$ & 1201 & 1.72\\
		NdFe$_{10}$Co(8f)$_{1.6}$As$_{0.4}$ & 1200 & 1.69\\
		NdFe$_{10}$Co(8f)$_{1.8}$Ge$_{0.2}$ & 1199 & 1.71\\
		NdFe$_{10}$Co(8f)$_{1.4}$As$_{0.6}$ & 1198 & 1.65\\
		NdFe$_{10}$Co(8f)$_{1.2}$As$_{0.8}$ & 1198 & 1.62\\
		NdFe$_{10}$Co(8f)$_{2.0}$           & 1198 & 1.76\\
		NdFe$_{10}$Co(8f)$_{1.6}$Ge$_{0.2}$As$_{0.2}$ & 1197 & 1.68\\
		NdFe$_{10}$Co(8f)As & 1197 & 1.59\\
		NdFe$_{10}$Co(8f)$_{1.6}$Ge$_{0.4}$ & 1195 & 1.68\\
		NdFe$_{10}$Co(8f)$_{0.8}$Ge$_{1.2}$ & 1195 & 1.56\\
		\hline
		NdFe$_{12}$ & 881 & 1.73 \\
		\hline
	\end{tabular}
\end{table}
All of the top ten systems contain
the maximal amount of dopants, namely, $x=1$ for list (A) and $x=2$ for list (B).
For list (A), doping with As and Ge give high scores than doping with Co.
For list (B), which could accommodate more dopant atoms, 
doping with Co was more advantageous. 
However, the system with the highest Curie temperature was obtained 
by co-doping with As and Co. 
From comparison of NdFe$_{10}$Co(8f)$_{1.8}$As$_{0.2}$ and 
NdFe$_{10}$Co(8f)$_{2}$,
the direct and indirect contributions of As
to the enhancement of $T_\mathrm{C}$ 
can be estimated as $-22$ and $+24$~K, respectively.
Arsenic enhances the magnetic couplings between surrounding transition metals,
and this effect is evidently slightly larger than that of the loss of the Co--Fe and Co--Co couplings
resulting from the substitution.
It is also noteworthy that substitution with Ge and As
can reduce the amount of Co,
which is an expensive element, 
without sacrificing the Curie temperature.

\section{Conclusion}
\label{S:Conclusion}
In this paper, we have discussed the effects of various dopants (K--Br)
on the magnetism of NdFe$_{12}$.
We first investigated doping with a single dopant at an infinitesimal concentration,
then extended the analysis to finite concentrations of selected dopants.
We have demonstrated the potential of As and Ge in enhancing
the Curie temperature and discussed the origin of this enhancement in terms of 
the interaction between the Fe 3d and $M$ 4p electrons.
These results were used to screen the dopants prior to further optimization
using a wider search space that allowed for
simultaneous doping.
We found that co-doping with As, Ge, and Co has the potential
to enhance the Curie temperature more efficiently than 
doping with Co alone. 
The results also indicate that doping with As and Ge can reduce
the amount of Co, which is a scarce element, 
without reducing the Curie temperature.

\section*{Acknowledgment}
This work was supported by 
a project (JPNP20019) commissioned
by the New Energy and Industrial Technology Development Organization (NEDO), 
the Elements Strategy Initiative Center for Magnetic Materials (ESICMM, 
Grant Number JPMXP0112101004), 
and the ``Program for Promoting Researches on the Supercomputer Fugaku''
(DPMSD) by MEXT.
The calculations were conducted in part using the facilities of the Supercomputer Center at
the Institute for Solid State Physics, University of Tokyo,
the supercomputer of the Academic Center for Computing and Media Studies (ACCMS), Kyoto University, 
and the supercomputer Fugaku provided by the RIKEN Center for Computational Science
through the HPCI System Research Project 
(Project ID: hp200125, hp210179).

\appendix
\section{Magnetic moment}
\subsection{NdFe$_{12-\delta}M_\delta\ (\delta \ll 1)$}
\begin{figure}[!t]
	\centering
	 \includegraphics[width=8cm]{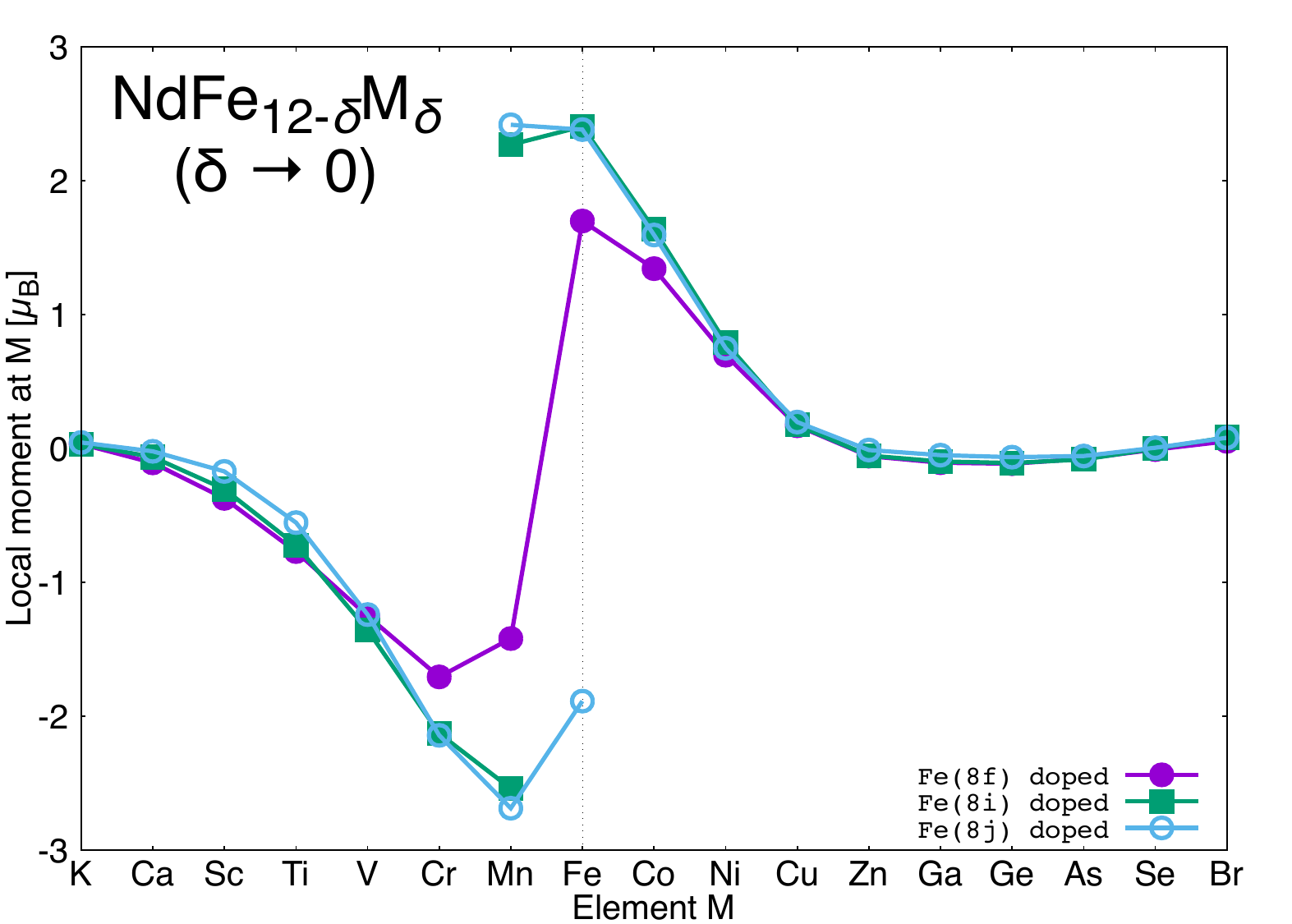}
	 \caption{\label{Fig:M0_moment}
	 Local magnetic moments of $M$ for Nd(Fe$_{12-\delta}M_\delta$)
	 in the dilute limit, $\delta \rightarrow 0$.}
\end{figure}
In this subsection, we present results for the magnetic moments
in NdFe$_{12-\delta}M_\delta\ (\delta \ll 1)$.
In Fig.~\ref{Fig:M0_moment}, the local magnetic moments of $M$
at the 8f, 8i, and 8j sites are shown.
Parallel and antiparallel solutions coexist
only in the cases of Fe(8i) doping with $M$ = Mn and 
Fe(8j) doping with $M$ = Fe and Mn.
Irrespective of the choice of doping site, 
the stable solution was antiparallel for $M$ = Mn and
parallel for $M$ = Fe.

It has been discussed in the context of magnetic impurity problems
that metastable states are more likely to exist
when the host has a small magnetic moment.\cite{Akai90}
The host is identical in our calculations,
but the existence of a metastable state
depends on the doping site.
The calculated local moments of Fe in NdFe$_{12}$ 
were 1.70 $\mu_B$ at the 8f site,
2.41 $\mu_B$ at the 8i site, and
2.38 $\mu_B$ at the 8j site.
Therefore, the Fe moments surrounding the 8f site were larger 
than those surrounding the other sites.
This leads to the non-existence of the metastable state
in the range of our calculations for 8f doping.

\begin{figure}[!t]
	\centering
	 \includegraphics[width=8cm]{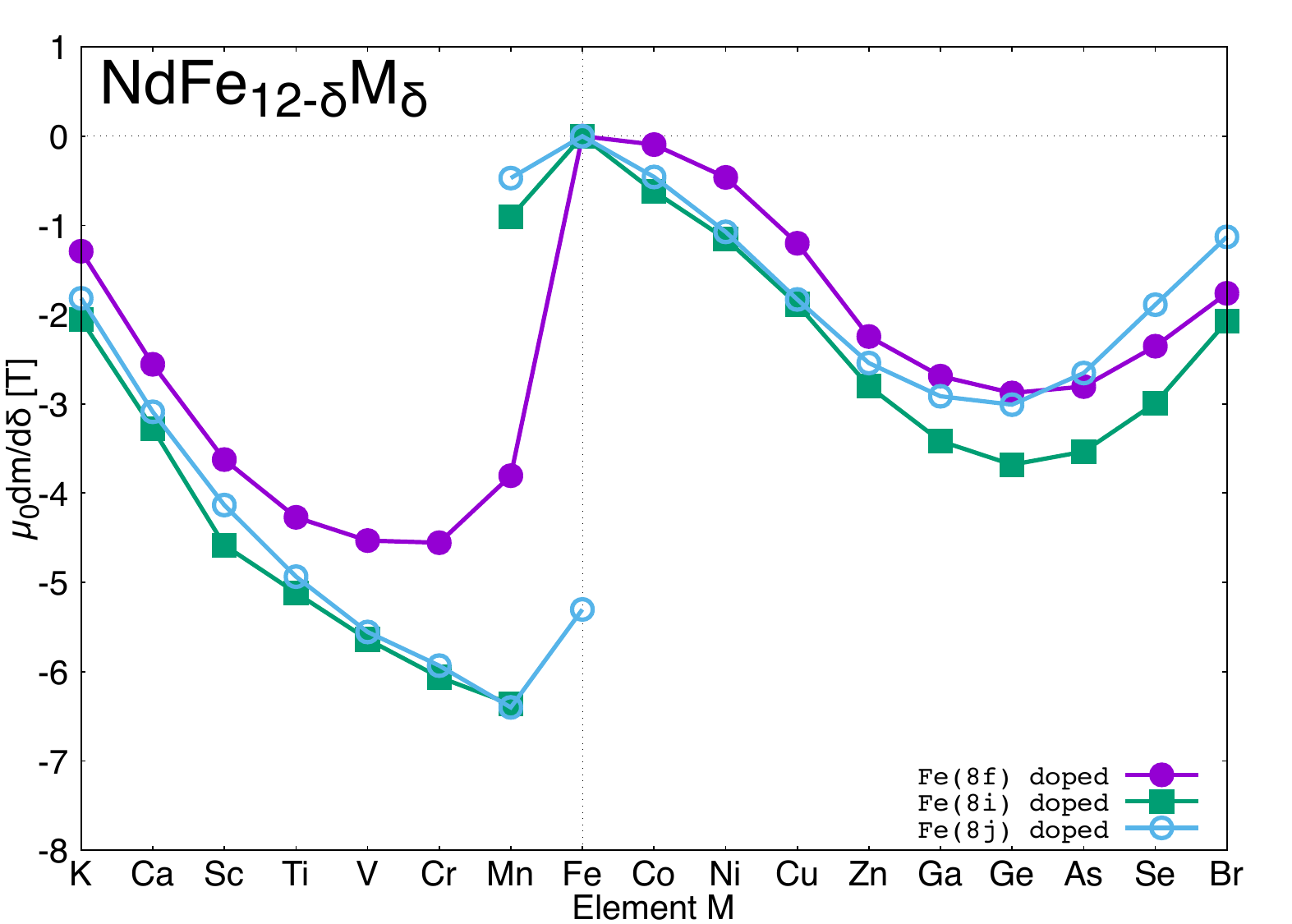}
	 \caption{\label{Fig:dM_dc}
	 Derivative of the magnetization for
	 Nd(Fe$_{12-\delta}M_\delta$)
	 with respect to $\delta$
	 in the dilute limit, $\delta \rightarrow 0$.}
\end{figure}

Let us next consider the rate of change of the magnetization $m$ 
with respect to the concentration.
Figure~\ref{Fig:dM_dc} shows
the derivative of the magnetization,
$\left.\frac{dm}{d\delta}\right|_{\delta=0}$,
as a function of the dopant $M$.
The magnetic moment decreases upon doping.
In the case of the elements to the left of the plot (K--Fe),
this is 
because the local moment of $M$ is antiparallel to the host.
In the case of those to the right of the plot (Mn--Br),
this is because the local moment of the replacing element
is smaller than
that of the replaced Fe,
which we call the direct contribution in 
DID analysis. 
However, this decrease is suppressed for $M$=Ge--Br,
even though the local moment of $M$ is almost zero.
This indicates that these elements have an indirect effect
on the magnetic moment which is proportional to the first order of
the concentration and counteracts the direct effect.

\newpage
\subsection{Finite amount of $M$}
\begin{figure*}[!t]
	\centering
	\includegraphics[width=35mm]{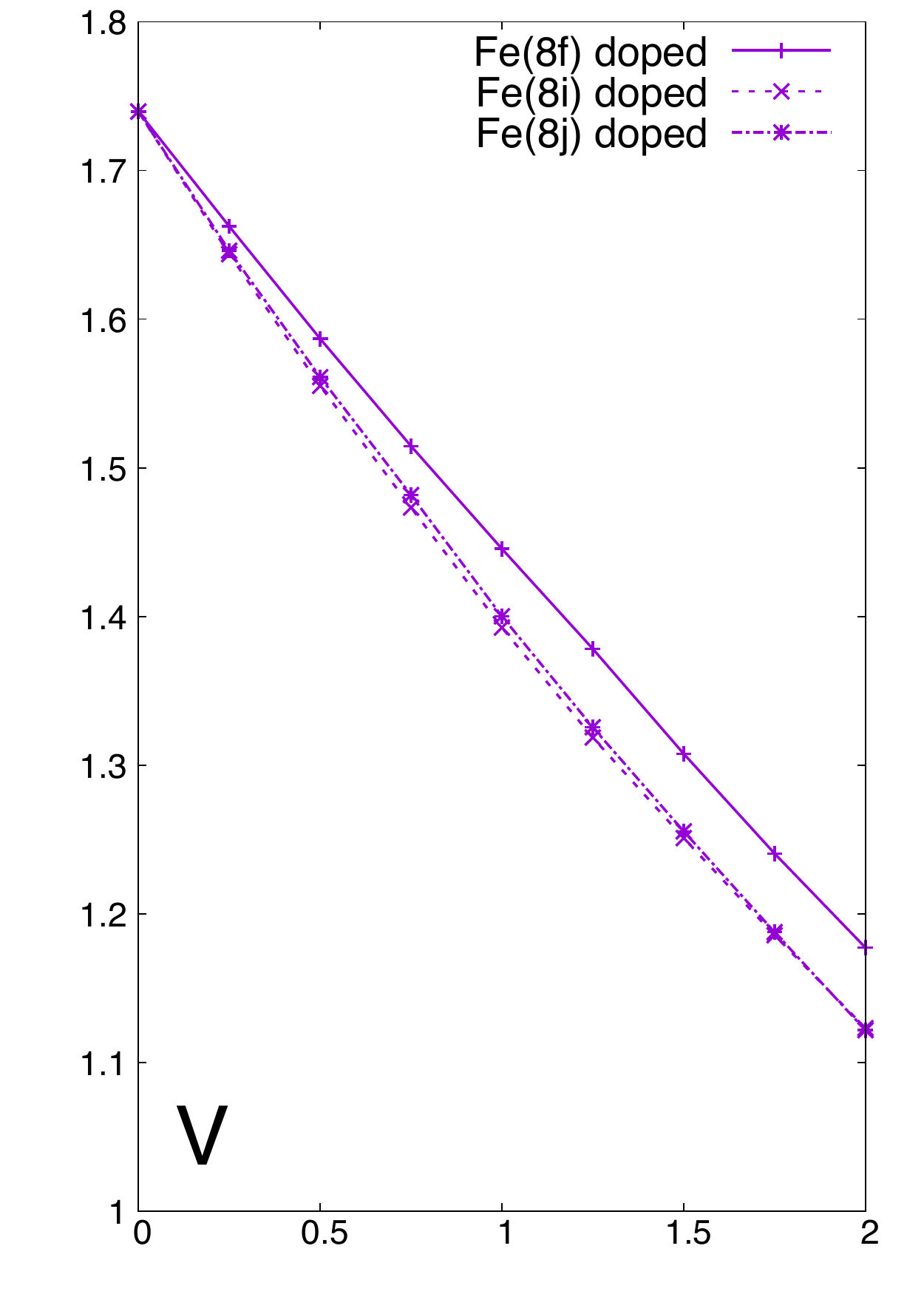}
	\includegraphics[width=35mm]{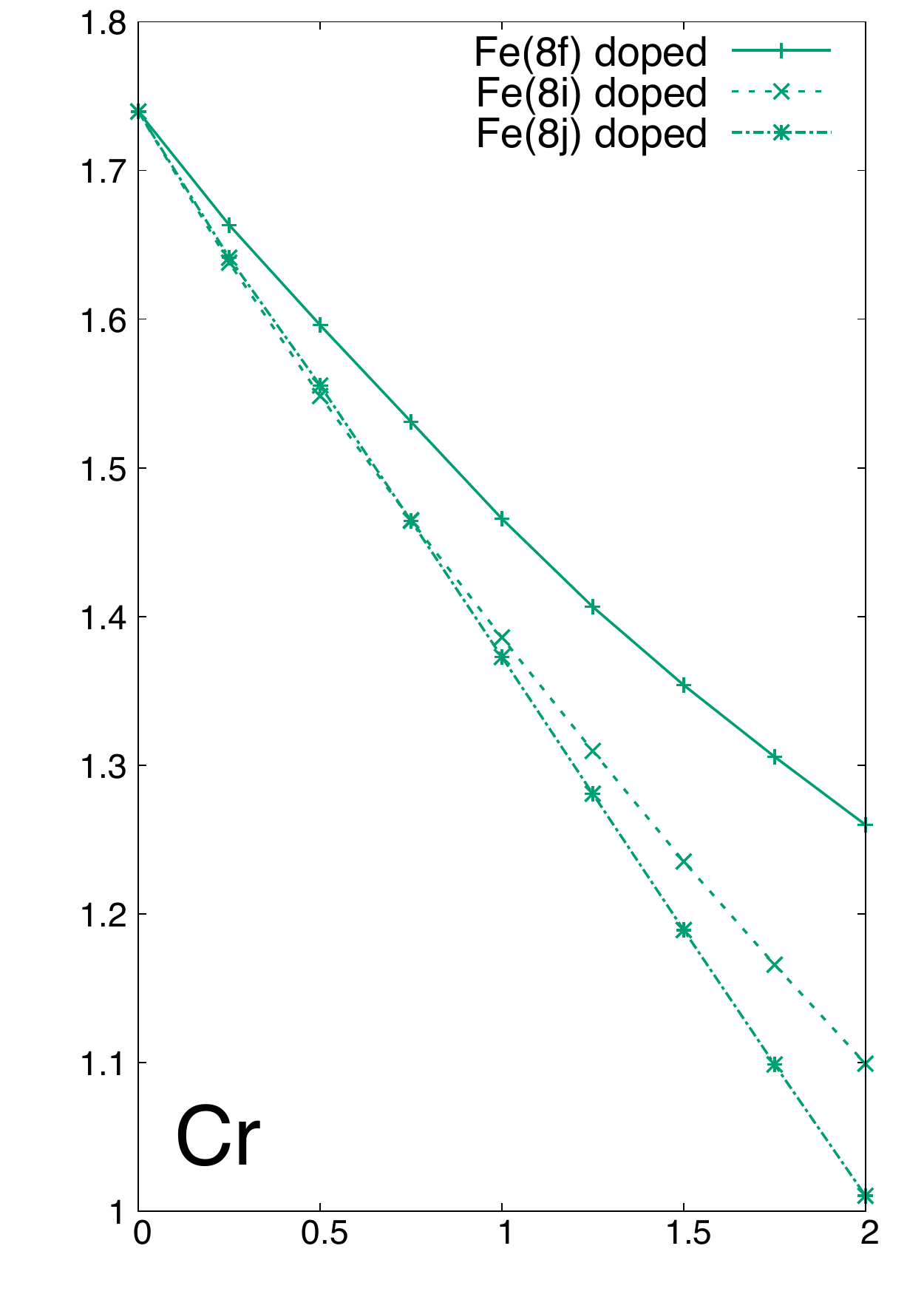}
    \includegraphics[width=35mm]{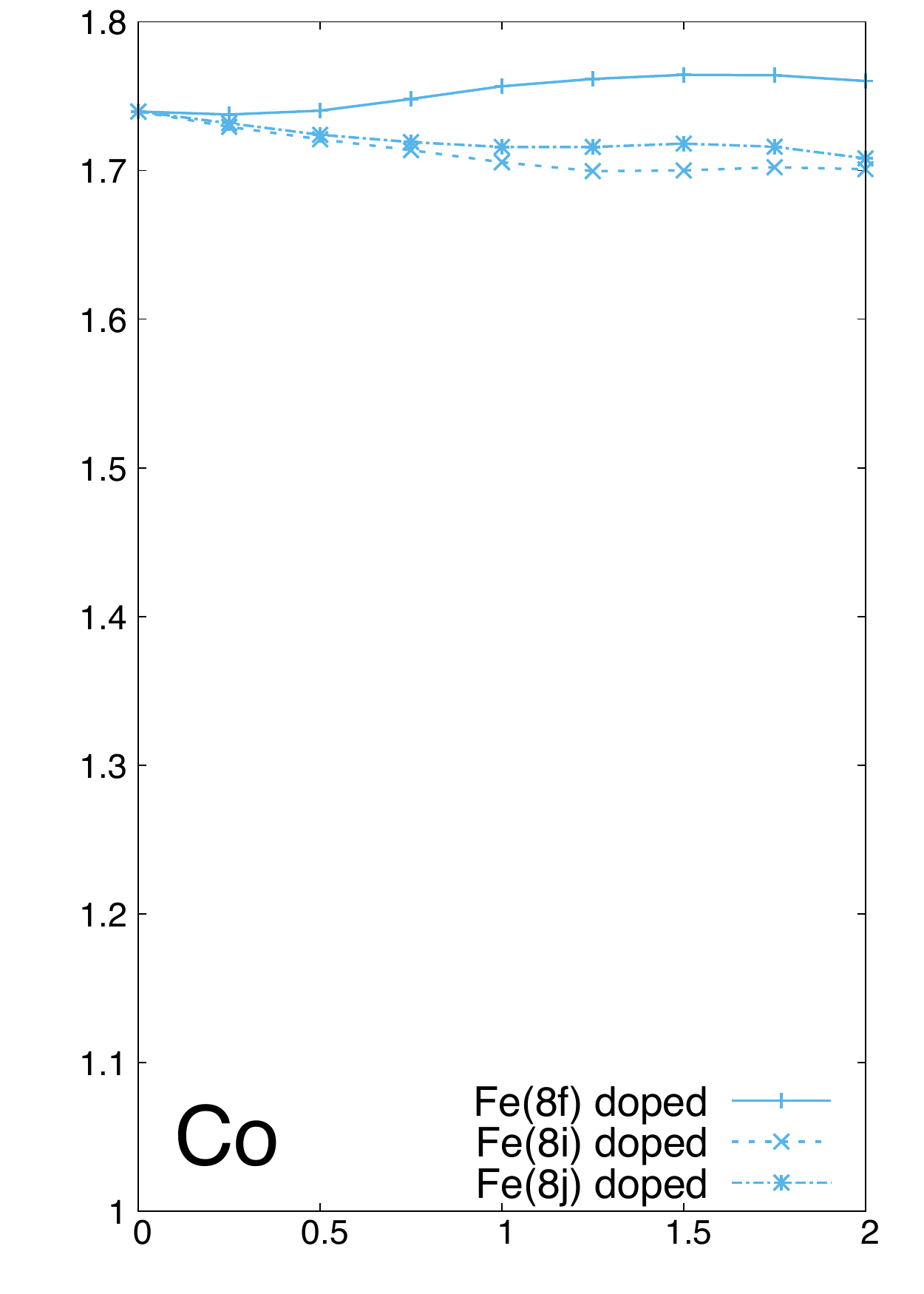}
    \\
	\includegraphics[width=35mm]{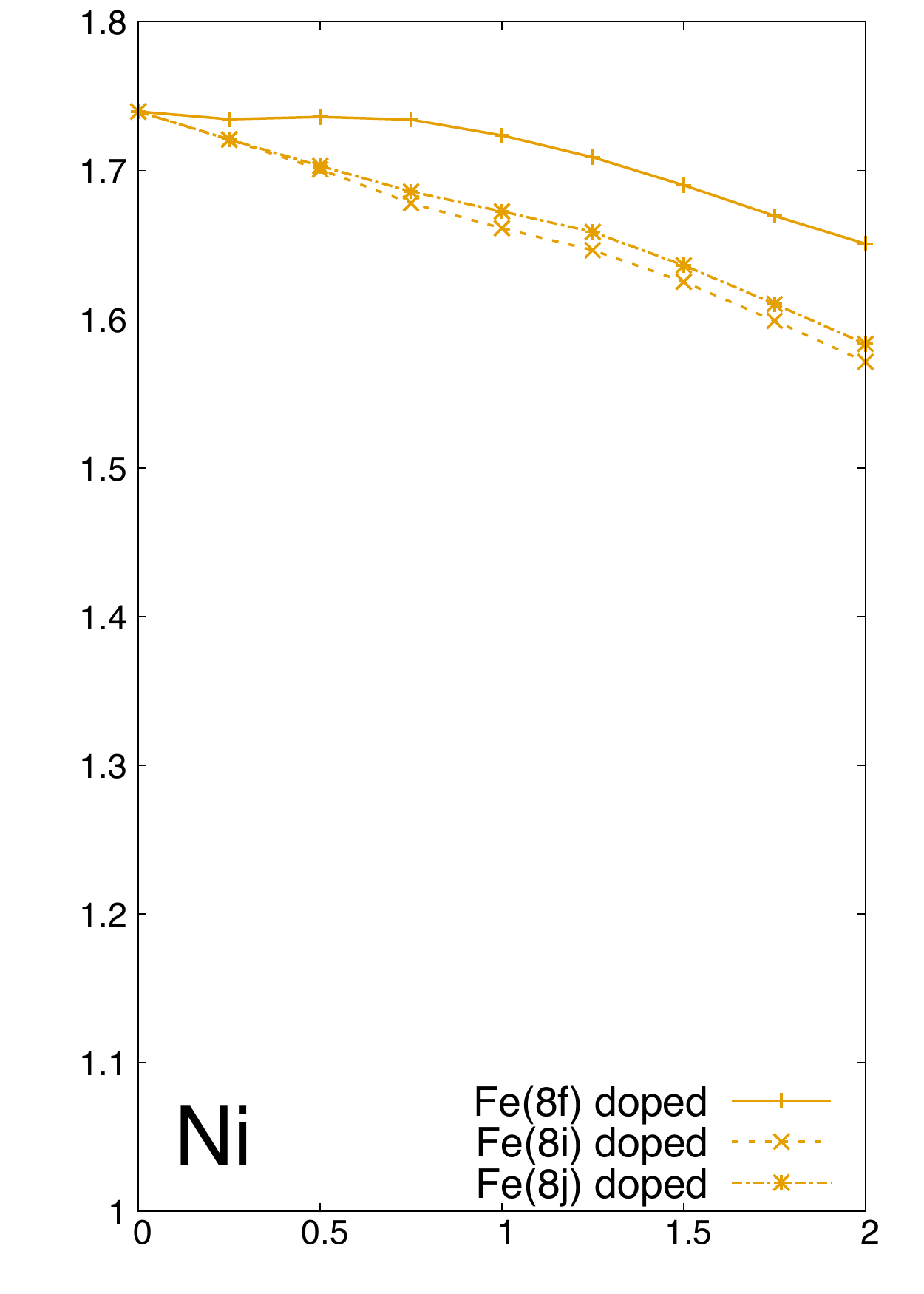}
	\includegraphics[width=35mm]{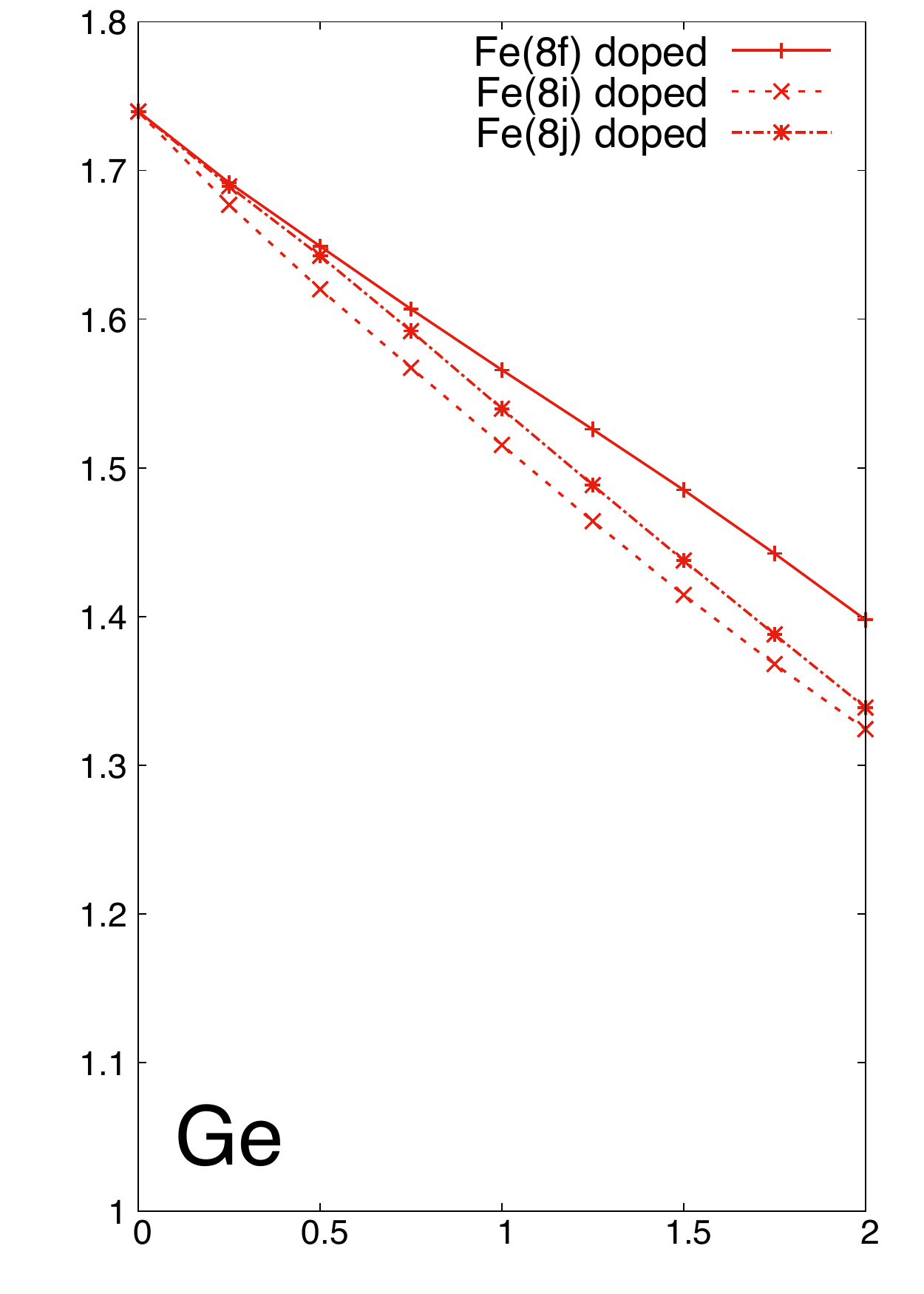}
	\includegraphics[width=35mm]{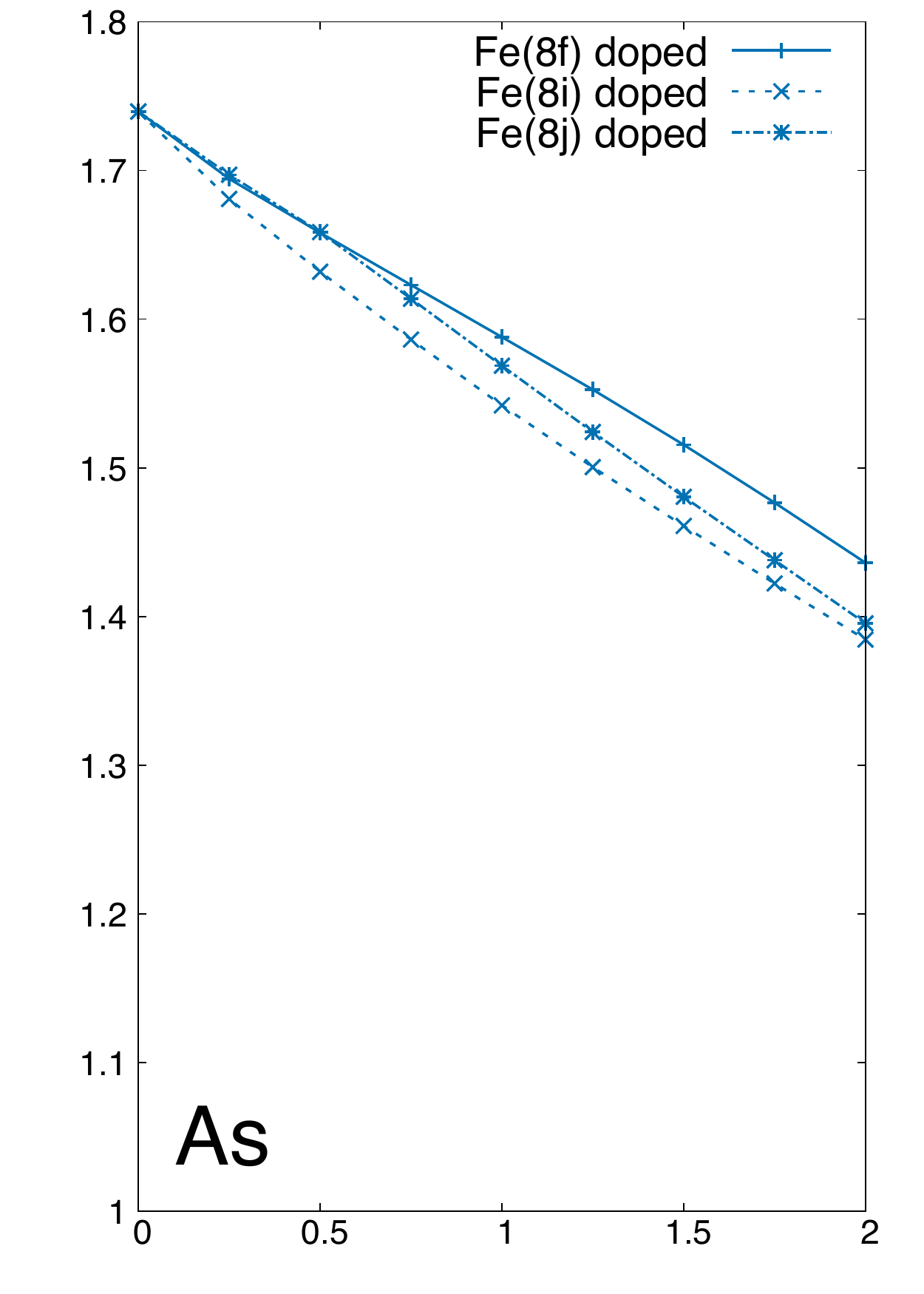}
	 \caption{\label{Fig:total_tesla_finite_conc}
	 Magnetization of Nd(Fe$_{12-x}$M$_x$) (M=V, Cr, Co, Ni, Ge, or As)
	 with finite concentrations $x$ in the range of $0 \leq x \leq 2$.}
\end{figure*}

Figure~\ref{Fig:total_tesla_finite_conc}
shows the results for the magnetization
of NdFe$_{12-x}M_x\ (0 \leq x \leq 2)$.
In the cases of $M$=V and Cr, the magnetic moment of $M$
is antiparallel to the host magnetization,
and the total magnetization sharply decreases
as the concentration of $M$ increases.
In the cases of $M$=Co and Ni, the magnetic moment of $M$
is parallel to the host, and the total magnetization is retained
at higher concentrations.
Because Ge and As are nonmagnetic, the decrease in the total 
magnetization is moderate compared with the cases of $M$=Ge and As.

\bibliography{nu}

\begin{thebibliography}{10}
\expandafter\ifx\csname url\endcsname\relax
  \def\url#1{\texttt{#1}}\fi
\expandafter\ifx\csname urlprefix\endcsname\relax\def\urlprefix{URL }\fi
\expandafter\ifx\csname href\endcsname\relax
  \def\href#1#2{#2} \def\path#1{#1}\fi

\bibitem{Hirayama15}
Y.~Hirayama, Y.~Takahashi, S.~Hirosawa, K.~Hono,
  \href{http://www.sciencedirect.com/science/article/pii/S1359646214004163}{{NdFe$_12$
  N$_x$ hard-magnetic compound with high magnetization and anisotropy field}},
  Scripta Materialia 95 (2015) 70--72.
\newline\urlprefix\url{http://www.sciencedirect.com/science/article/pii/S1359646214004163}

\bibitem{Hirayama15b}
Y.~Hirayama, T.~Miyake, K.~Hono, Rare-earth lean hard magnet compound ndfe12n,
  JOM 67~(6) (2015) 1344--1349.

\bibitem{Ohashi87}
K.~Ohashi, T.~Yokoyama, R.~Osugi, Y.~Tawara, {The magnetic and structural
  properties of R-Ti-Fe ternary compounds}, IEEE transactions on magnetics
  23~(5) (1987) 3101--3103.

\bibitem{Ohashi88}
K.~Ohashi, Y.~Tawara, R.~Osugi, J.~Sakurai, Y.~Komura,
  \href{http://www.sciencedirect.com/science/article/pii/0022508888900203}{Identification
  of the intermetallic compound consisting of sm, ti, fe}, Journal of the Less
  Common Metals 139~(2) (1988) L1--L5.
\newline\urlprefix\url{http://www.sciencedirect.com/science/article/pii/0022508888900203}

\bibitem{Harashima16}
Y.~Harashima, K.~Terakura, H.~Kino, S.~Ishibashi, T.~Miyake, {First-principles
  study on stability and magnetism of NdFe$_{11}$$M$ and NdFe$_{11}$$M$N for M=
  Ti, V, Cr, Mn, Fe, Co, Ni, Cu, Zn}, arXiv preprint arXiv:1609.07227 (2016).

\bibitem{Hirayama17}
Y.~Hirayama, Y.~K. Takahashi, S.~Hirosawa, K.~Hono, Intrinsic hard magnetic
  properties of {Sm(Fe$_{1-x}$Co$_x$)$_{12}$} compound with the {ThMn$_{12}$}
  structure, Scripta Materialia 138 (2017) 62--65.

\bibitem{Fukazawa18}
T.~Fukazawa, H.~Akai, Y.~Harashima, T.~Miyake, {First-principles study of
  intersite magnetic couplings and Curie temperature in RFe$_{12-x}$Cr$_x$ (R =
  Y, Nd, Sm)}, Journal of Physical Society of Japan 87~(4) (2018) 044706.

\bibitem{Schoenhoebel19}
A.~Sch{\"o}nh{\"o}bel, R.~Madugundo, O.~Y. Vekilova, O.~Eriksson, H.~C. Herper,
  J.~Barandiar{\'a}n, G.~Hadjipanayis, Intrinsic magnetic properties of smfe12-
  xvx alloys with reduced v-concentration, Journal of Alloys and Compounds 786
  (2019) 969--974.

\bibitem{Fukazawa19c}
T.~Fukazawa, Y.~Harashima, Z.~Hou, T.~Miyake, {Bayesian optimization of
  chemical composition: A comprehensive framework and its application to
  $R$Fe$_{12}$-type magnet compounds}, Physical Review Materials 3~(5) (2019)
  053807.

\bibitem{Hohenberg64}
P.~Hohenberg, W.~Kohn, Inhomogeneous electron gas, Physical Review 136~(3B)
  (1964) B864.

\bibitem{Kohn65}
W.~Kohn, L.~J. Sham, Self-consistent equations including exchange and
  correlation effects, Physical Review 140~(4A) (1965) A1133.

\bibitem{Korringa47}
J.~Korringa, On the calculation of the energy of a bloch wave in a metal,
  Physica 13~(6-7) (1947) 392--400.

\bibitem{Kohn54}
W.~Kohn, N.~Rostoker,
  \href{http://link.aps.org/doi/10.1103/PhysRev.94.1111}{Solution of the
  schr\"odinger equation in periodic lattices with an application to metallic
  lithium}, Phys. Rev. 94 (1954) 1111--1120.
\newblock \href {https://doi.org/10.1103/PhysRev.94.1111}
  {\path{doi:10.1103/PhysRev.94.1111}}.
\newline\urlprefix\url{http://link.aps.org/doi/10.1103/PhysRev.94.1111}

\bibitem{Perdew81}
J.~P. Perdew, A.~Zunger,
  \href{http://link.aps.org/doi/10.1103/PhysRevB.23.5048}{Self-interaction
  correction to density-functional approximations for many-electron systems},
  Phys. Rev. B 23 (1981) 5048--5079.
\newblock \href {https://doi.org/10.1103/PhysRevB.23.5048}
  {\path{doi:10.1103/PhysRevB.23.5048}}.
\newline\urlprefix\url{http://link.aps.org/doi/10.1103/PhysRevB.23.5048}

\bibitem{Soven67}
P.~Soven, Coherent-potential model of substitutional disordered alloys,
  Physical Review 156~(3) (1967) 809.

\bibitem{Soven70}
P.~Soven, Application of the coherent potential approximation to a system of
  muffin-tin potentials, Physical Review B 2~(12) (1970) 4715.

\bibitem{Shiba71}
H.~Shiba, A reformulation of the coherent potential approximation and its
  applications, Progress of Theoretical Physics 46~(1) (1971) 77--94.

\bibitem{Harashima15g}
Y.~Harashima, K.~Terakura, H.~Kino, S.~Ishibashi, T.~Miyake,
  \href{http://journals.jps.jp/doi/abs/10.7566/JPSCP.5.011021}{First-principles
  study of structural and magnetic properties of r (fe, ti) 12 and r (fe, ti)
  12n (r= nd, sm, y)}, in: Proceedings of Computational Science Workshop 2014
  (CSW2014), Vol.~5 of JPS Conference Proceedings, 2015, p. 1021.
\newline\urlprefix\url{http://journals.jps.jp/doi/abs/10.7566/JPSCP.5.011021}

\bibitem{Liechtenstein87}
A.~I. Liechtenstein, M.~Katsnelson, V.~Antropov, V.~Gubanov, Local spin density
  functional approach to the theory of exchange interactions in ferromagnetic
  metals and alloys, Journal of Magnetism and Magnetic Materials 67~(1) (1987)
  65--74.

\bibitem{Fukazawa19b}
T.~Fukazawa, H.~Akai, Y.~Harashima, T.~Miyake,
  \href{https://ieeexplore.ieee.org/document/8653984}{{Curie temperature of
  Sm$_2$Fe$_{17}$ and Nd$_2$Fe$_{14}$B: a first-principles study}}, IEEE
  Transaction on Magnetics (in press).
\newline\urlprefix\url{https://ieeexplore.ieee.org/document/8653984}

\bibitem{Ueno16}
T.~Ueno, T.~D. Rhone, Z.~Hou, T.~Mizoguchi, K.~Tsuda, {COMBO: An efficient
  Bayesian optimization library for materials science}, Materials discovery 4
  (2016) 18--21.

\bibitem{COMBO}
\href{https://github.com/tsudalab/combo}{{COMmon Bayesian Optimization Library
  (COMBO)}}, \url{https://github.com/tsudalab/combo}.
\newline\urlprefix\url{https://github.com/tsudalab/combo}

\bibitem{Kanamori90}
J.~Kanamori, Interplay between electronic structure and correlation through the
  sd mixing in transition metal systems, Progress of Theoretical Physics
  Supplement 101 (1990) 1--10.

\bibitem{Ogura11}
M.~Ogura, H.~Akai, J.~Kanamori, {Enhancement of Magnetism of Fe by Cr and V},
  Journal of the Physical Society of Japan 80~(10) (2011) 104711.

\bibitem{Harashima15e}
Y.~Harashima, K.~Terakura, H.~Kino, S.~Ishibashi, T.~Miyake,
  \href{http://link.aps.org/doi/10.1103/PhysRevB.92.184426}{Nitrogen as the
  best interstitial dopant among {$X$=B, C, N, O, and F} for strong permanent
  magnet {${\mathrm{NdFe}}_{11}\mathrm{Ti}X$}: First-principles study}, Phys.
  Rev. B 92 (2015) 184426.
\newblock \href {https://doi.org/10.1103/PhysRevB.92.184426}
  {\path{doi:10.1103/PhysRevB.92.184426}}.
\newline\urlprefix\url{http://link.aps.org/doi/10.1103/PhysRevB.92.184426}

\bibitem{Akai90}
H.~Akai, M.~Akai, S.~Bl{\"u}gel, B.~Drittler, H.~Ebert, K.~Terakura, R.~Zeller,
  P.~Dederichs,
  \href{http://ptps.oxfordjournals.org/content/101/11.short}{Theory of
  hyperfine interactions in metals}, Progress of Theoretical Physics Supplement
  101 (1990) 11--77.
\newline\urlprefix\url{http://ptps.oxfordjournals.org/content/101/11.short}

\end{thebibliography}
\end{document}